\documentclass[a4paper,11pt]{article}
\usepackage{etoolbox}
\usepackage{dcolumn}
\usepackage{bm}
\usepackage{graphics}
\usepackage{epstopdf}
\usepackage{tikz,xcolor}
\usepackage{dirtytalk}
\usepackage{graphicx} 
\usepackage{blindtext}
\usepackage{amsmath}
\usepackage{mathrsfs}
\usepackage{amsfonts}
\usepackage{amssymb}
\usepackage{mathtools,halloweenmath}
\usepackage[%
colorlinks=true,
urlcolor=blue,
linkcolor=red,
citecolor=blue
]{hyperref}
\usepackage{graphicx}
\usepackage{wrapfig}
\usepackage{color}
\usepackage{subcaption}
\usepackage{dsfont}
\usepackage{dutchcal}
\usepackage{verbatim}
\usepackage{float}
\usepackage{lipsum}
\usepackage{cite}
\usepackage[title]{appendix}
\usepackage{physics}

\definecolor{lime}{HTML}{A6CE39}
\DeclareRobustCommand{\orcidicon}{%
	\begin{tikzpicture}
	\draw[lime, fill=lime] (0,0) 
	circle [radius=0.16] 
	node[white] {{\fontfamily{qag}\selectfont \tiny ID}};
	\draw[white, fill=white] (-0.0625,0.095) 
	circle [radius=0.007];
	\end{tikzpicture}
	\hspace{-2mm}
}

\foreach \x in {A, ..., Z}{%
	\expandafter\xdef\csname orcid\x\endcsname{\noexpand\href{https://orcid.org/\csname orcidauthor\x\endcsname}{\noexpand\orcidicon}}
}

\begin{document}
\title{\textbf {Enhancement of an Unruh-DeWitt battery performance through quadratic environmental coupling}}
\vskip 1cm
\author{
{\bf {\normalsize Arnab Mukherjee}\orcidA{}\thanks{arnab.mukherjee@bose.res.in}},
{\bf {\normalsize Sunandan Gangopadhyay}\orcidB{}},
{\bf {\normalsize A. S. Majumdar}}
\\
{\normalsize Department of Astrophysics and High Energy Physics},\\
{\normalsize S.N. Bose National Centre for Basic Sciences},\\
{\normalsize JD Block, Sector III, Salt Lake, Kolkata 700106, India}\\
}
\maketitle

\begin{abstract}
	\noindent We investigate relativistic effects on the performance of a quantum battery  in an open quantum framework. We consider an Unruh-DeWitt detector driven by a coherent classical pulse as a quantum battery that is interacting with a massless scalar field through a quadratic coupling. The battery follows a trajectory composed of uniform acceleration along one direction, combined with constant four-velocity components in the orthogonal plane to the acceleration. Accelerated motion degrades the performance of the quantum battery rapidly in the absence of the orthogonal velocity component. We first derive the Lindblad equation for quadratic coupling in detail. We then show that the quadratic scalar field coupling enhances coherence and stability in the presence of orthogonal velocity. We observe that decoherence is mitigated significantly, resulting in remarkable improvement in the battery capacity and efficiency compared to the case of the usual linear field coupling. This opens up the possibility of nonlinear environmental coupling enabling stored energy to be retained over longer durations, leading to more efficient operation of quantum devices. 
\end{abstract}
\section{Introduction}\label{sec:Intro}
\noindent The field of quantum thermodynamics has emerged as a highly active area of research due to the recent advancements in quantum technologies \cite{kosloff2013quantum, Vinjanampathy2016, Bhattacharjee2021, RMP_QB_2024},
enabling probe of length scales of the order of nano-meters. The advent of technological miniaturization and the keen urge for energy manipulation have created an upsurge of interest in the theoretical and experimental study of an alternative and efficient energy storage device, known as a quantum battery \cite{Alicki2013, Campaioli2018}. Quantum batteries are models of open quantum systems interacting with a quantum field acting as an environment. 

Using the non-classical properties of quantum systems, such as quantum superposition, coherence, entanglement, and many-body collective behaviours, quantum batteries are able to perform faster and more efficient charging processes than their classical counterparts \cite{Campisi2011, Horodecki2013, Goold2016, Campaioli2018, Binder2015, Binder2017, Farina2019, Zhang_YY2019, Carrasco2022, Dario_2022_1, Rodriguez2023, Gemme2023, Gemme2024, Song2024}. Quantum batteries can be realised though various systems such as fluorescent organic molecules 
embedded in a microcavity \cite{Quach2022}, IBM quantum chips \cite{Gemme2022}, transmon qubits \cite{Hu_CK2022}, quantum dots \cite{Wenniger2023}, two-level system coupled to a photonic waveguide \cite{Lu_2024, Tirone_2024_1}, based on the Sachdev-Ye-Kitaev (SYK) model \cite{Dario_2020, Dario_JHEP_2020, Dario_2022}, and based on the collision model \cite{Shaghaghi_2022, Dario_2023}. The interaction of a quantum battery with its surroundings results in the coherence of the battery leaking to the surroundings. Due to this decoherence effect, the charging and discharging performance of quantum batteries is frequently affected \cite{Farina2019, Carrega2020, Tabesh2020}.

The study of quantum batteries in the relativistic framework is
inspired from the fundamental perspective of understanding quantum effects in the
relativistic regime, and no less from the current technological outlook
of satellite based quantum networks such as the quantum internet \cite{Lu_2022, Ribezzo_2023}. From the seminal works \cite{Fulling, Davies1975, Unruh1}, it is well known that uniformly accelerated systems exhibit the Unruh effect. In relativistic framework,  degradation of quantum resources such as coherence and entanglement \cite{Bruschi2014, Matsas, Y_M_Huang2019, Z_Zhao2020, Du2021, Liu2021, Benatti_2004, B_L_Hu2012, Moustos2017, Chatterjee2017, Mukherjee_PRA_2024, Sokolov2020} occurs.
Recently,  it has been noticed that studying  quantum heat engines in the relativistic framework leads to some exciting results in the context of work extraction and the efficiency \cite{arias2018unruh, Mukherjee2022}.

In the context of quantum batteries, \textit{ergotropy, battery capacity}, and \textit{charging efficiency} are the fundamental quantities which govern the performance of the battery \cite{Allahverdyan2004, M_N_Bera2020, Mir2023, Mojaveri2023, Fischer_2024, Tirone_2024}.
Work extraction from a quantum battery has been studied in the relativistic framework \cite{Hao2023}, and it was shown that work extraction of a quantum battery can be degraded due to its accelerated motion.  However, such
degradation can be inhibited by moving the battery along a trajectory characterized by the combination of a linear accelerated motion along with a component of the four-velocity in an orthogonal direction\cite{Abdolrahimi2014, Liu2021}. Further, the charging efficiency of a quantum battery can be enhanced by moving the battery and the charger at a nonrelativistic velocity \cite{Mojaveri2023}. In a recent study, the charging process of a quantum battery is studied in a curve background \cite{Tian_2024}.

The above investigations on quantum batteries have been based on the framework of
the simplest model of a quantum detector, {\it viz.},  a two-level atom or
qubit, usually known as the Unruh-De Witt detector \cite{DeWitt1979, Unruh1984} interacting linearly with a scalar field, resembling the interaction of an atomic dipole coupled with the electromagnetic field \cite{hu2012geometric,Martinez2016_1}. In the open quantum framework, a plethora of works have been investigated using a single \cite{Benatti_2004, yu2008understanding, hu2012geometric, jin2014dynamical,liu2021relativistic} and two Unruh-DeWitt detectors \cite{Benatti_2004,zhang2007unruh,hu2015entanglement,yang2016entanglement,cheng2018entanglement,she2019entanglement,zhou2021entanglement, chen2022entanglement} in diverse physical scenarios. A natural extension of these genres of studies is to employ quadratic coupling of the quantum battery with the scalar field 
environment  \cite{Alicki2013}. The primary motivation for studying quadratic interactions stems from the fact that quadratic interaction is analogous to the phenomenon of the \textit{two-photon absorption (TPA) process}, which has fundamental importance in the field of quantum optics \cite{gilles1993tp}. From the reference \cite{Wu2023}, we also find that the entanglement between two Unruh-DeWitt detectors with the quadratic field coupling scenario does not decrease monotonically with the increase in acceleration, in contrast to the linear field coupling scenario. Quadratic field couplings for Unruh-DeWitt detectors have been considered earlier 
\cite{hinton1984particle, Hummer2016, Sachs2017, Gray2018}, leading to the possibility of  enhanced entanglement harvesting. It may also be relevant to note here that
environmental engineering for control of open quantum systems forms a
key focus of current research in quantum devices \cite{Koch_2016, Uchiyama_2018}. Recently, it has been shown that the non-linear nature of the TPA process opens up the possibility of exploiting quantum resources, such as entanglement, to enhance measurement precision and access new information about ultrafast molecular dynamics. In addition, this non-linear interaction provides a promising route for establishing the metrological advantages of nonclassical squeezed light sources in precision measurements of TPA cross sections \cite{munoz2021quantum}. Quadratic light-matter interaction has also been studied from the perspective of waveguide QED framework \cite{alushi2023waveguide}. Previous works show that  superconducting qubits nonlinearly coupled to quantum microwave
resonators \cite{felicetti2018tp, felicetti2018ultra},  or nanomechanical oscillators \cite{zhou2006nonlinear}.

In the present study our aim is to investigate the effect of the accelerated motion  and the vacuum fluctuations of the massless scalar field on the fundamental parameters associated with the quantum battery by considering a quadratic interaction between the quantum battery and the scalar field environment. Due to the vacuum fluctuations during the charging time interval of the classical pulse, evolution of the quantum battery takes place. Combining the linear acceleration and two components of the four-velocity in the plane orthogonal to the acceleration, here we  propose a general form of the trajectory used in \cite{Liu2021, Hao2023}. We also
investigate whether the performance of the battery can be enhanced by tuning the velocity components of the battery. In order to evaluate the performance
of a quantum battery in relativistic settings, we evaluate the relevant quantities of battery ergodicity, capacity and efficiency. It may be noted that the battery capacity represents the ability of a quantum device to
store and supply energy without depending upon the temporary energy level
of the system \cite{Mir2023, Tirone_2024}. Our results display the possibility of significant enhancement of battery capacity and efficiency through quadratic
environmental coupling in the relativistic arena. 

The paper is organised as follows: In section \ref{sec:Model}, we briefly describe the model of a relativistic quantum battery and the fundamental quantities related with it. In section \ref{sec:GKSL}, we present a detailed derivation of the GKSL master equation for the scenario while the atom is quadratically coupled with the massless quantum scalar field. Section \ref{sec:Env_effect} describes the effect of the vacuum fluctuation on the battery. In section \ref{sec:Dyn_effect}, we study the dynamics of the quantum battery. In section \ref{sec:Rel_effect}, relativistic effects on the quantum battery parameters are calculated. We 
present our results on the battery parameters for different cases with
extensive discussions in section \ref{sec:Results}. Finally, we conclude with a summary of our results in section \ref{sec:Conclusion}. Throughout the paper, we take $\hbar=c=k_B=1$, where $k_B$ is the Boltzmann constant.
\section{Unruh-DeWitt battery and its parameters}\label{sec:Model}
\noindent In this section, we briefly describe the model of a relativistic quantum battery using a single Unruh-DeWitt (UDW) detector and some of the important parameters which are associated with its performance. 

We consider that the UDW detector is modeled as a two level system with two energy levels $\vert g\rangle$ and $\vert e\rangle$. The corresponding energy values of each level are $-\omega_{0}/2$ and $+\omega_{0}/2$ respectively. Initially we consider that the UDW detector is in its ground state $\vert g\rangle$. The initial Hamiltonian of the UDW detector is given by $H_{D}=\frac{1}{2}(\omega_{0}\sigma_z)$, where $\sigma_z=\vert e\rangle\langle e\vert-\vert g\rangle\langle g\vert$ is the third Pauli matrix. Now applying a time-dependent classical pulse to the UDW detector, the total Hamiltonian of the quantum battery becomes \cite{Gemme2022}
\begin{equation}
H''_{B}=\frac{1}{2}(\omega_{0}\sigma_z)+\alpha f(t) (\sigma_{+}+\sigma_{-})
\end{equation}
where $f(t)=f_0\cos{(\omega t)}$ is the time dependent amplitude of the classical pulse, and $\alpha$ is the effective coupling strength between the UDW detector and the classical drive. Here we set $f_0=1$ for $0< t\leq \tau$ where $\tau$ is the total charging time \cite{Hao2023}. As we consider $\omega_0>>\alpha$, therefore, in order to charge up the quantum battery, the driving frequency $\omega$ must be resonant with the initial energy gap of the UDW detector $\omega_0$ \cite{Carrega2020}, that is, $\omega=\omega_{0}$. This is called the resonance condition. Under the resonance condition and small coupling, the total Hamiltonian of the quantum battery in the rotating wave approximation can be written as \cite{scully1997quantum}
\begin{equation}\label{rwa}
H'_{B}=\frac{1}{2}(\omega_{0}\sigma_z)+\frac{\alpha}{2} (\sigma_{+}e^{-i\omega_{0} \tau}+\sigma_{-}e^{i\omega_{0} \tau})~.
\end{equation}
To get rid of this time dependency and make a simpler form of the Hamiltonian we take a rotating frame transformation \cite{Gemme2022}
\begin{equation}\label{rft}
    H_{B}=UH'_{B}U^{\dagger}-iU\frac{d}{d\tau}U^{\dagger}
\end{equation}
where $U$ is the unitary operator. Taking $U=e^{iH_D \tau}=e^{i\frac{\omega_{0}}{2}\sigma_z \tau}$ and putting eq.~\eqref{rwa} in eq.~\eqref{rft}, the total Hamiltonian of the quantum battery becomes
\begin{equation}
H_{B}=\frac{\alpha}{2}(\sigma_{+} +\sigma_{-})
\end{equation}
where $\sigma_{+}=\vert e\rangle\langle g\vert$ and $\sigma_{-}=\vert g\rangle\langle e\vert$ are the raising and lowering operators, respectively.

During the charging process, the state of the detector also evolve from $\rho_0$ to $\rho(\tau)$ such that 
$\rho(\tau)=U(\tau)\rho_0 U^{\dagger}(\tau)$ where $U(\tau)$ is the cyclical unitary operator. Hence, the net amount of energy stored in a quantum battery at charging time $\tau$ is given by \cite{Mojaveri2023}
\begin{equation}
\Delta E_{B}= Tr[\rho(\tau) H_{B} (\tau)] - Tr[\rho_{0} H_{B} (\tau)]~.
\end{equation}
In order to estimate the performance of the quantum batteries, we need to study some key parameters of these objects. An important parameter in the study of quantum batteries is the ergotropy \cite{Allahverdyan2004}, which is the maximum amount of energy extracted from a given quantum state $\rho(\tau)$ through a cyclic unitary operation. This is given by \cite{Alicki2013}
\begin{align}
\mathcal{E}&=Tr[\rho(\tau) H_{B}(\tau)] - \mathop{\text{min}}_{\{U\}}Tr[U(\tau)\rho(\tau) U^{\dagger}(\tau) H_{B}(\tau)]\,\nonumber\\
&=Tr[\rho(\tau) H_{B}(\tau)] -Tr[U_{\sigma}(\tau)\rho(\tau) U^{\dagger}_{\sigma}(\tau) H_{B}(\tau)]
\end{align}
where the minimization is taken over all possible unitary transformations acting locally on such system. The minimal unitary transformation $U_{\sigma}$ corresponds to the fact that the states $\sigma_{B}\equiv U_{\sigma}\rho(\tau) U^{\dagger}_{\sigma}=\displaystyle\sum_{i=1}^{2} s_i\vert \varepsilon_i\rangle\langle\varepsilon_i\vert$ are passive states. Here, $\vert \varepsilon_i\rangle$ is the energy eigenstate of $H_{B}(\tau)$ and {$s_i$} are the eigenvalues of $\rho(\tau)$ are arranged in the decreasing order.

It is also relevant to consider  another quantity known as antiergotropy, which is defined as the minimum amount of energy extractable from the quantum state, given by \cite{Mir2023}
\begin{align}
\mathcal{A}&=Tr[\rho(\tau) H_{B}(\tau)] - \mathop{\text{max}}_{\{U\}}Tr[U(\tau)\rho(\tau) U^{\dagger}(\tau) H_{B}(\tau)]\,\nonumber\\
&=Tr[\rho(\tau) H_{B}(\tau)] -Tr[U_{\pi}(\tau)\rho(\tau) U^{\dagger}_{\pi}(\tau) H_{B}(\tau)]
\end{align}
where the maximization is taken over all possible unitary transformations and the maximal unitary transformation $U_\pi$ corresponds to the fact that the states $\pi_{B}\equiv U_\pi\rho(\tau) U^{\dagger}_\pi=\displaystyle\sum_{i=1}^{2} s_i\vert \varepsilon_i\rangle\langle\varepsilon_i\vert$ are active states.
Here, the eigenvalues of $\rho(\tau)$ are arranged in the increasing order.

In the study of quantum batteries, a figure of merit representing the potential of a quantum system to store and transfer energy  is the battery capacity \cite{M_N_Bera2020, Mir2023}. This quantity represents the amount of work that a quantum system can transfer during any thermodynamic cycle keeping the evolution of the battery to be unitary, and is given by the difference of the ergotropy and the antiergotropy. Using the definition of ergotropy and  antiergotropy, the quantum battery capacity can be written as \cite{Mir2023}
\begin{equation}
\mathcal{C}=\mathcal{E}-\mathcal{A}=Tr[\pi_{B}H_{B}(\tau)]-Tr[\sigma_{B}H_{B}(\tau)]\,.
\end{equation}
With the help of the ergotropy $\mathcal{E}$ and the net energy in the charging process $\Delta E_{B}$, the efficiency $\eta$ of the quantum battery is defined as \cite{Mojaveri2023}
\begin{equation}
\eta=\frac{\mathcal{E}}{\Delta E_{B}}\,.
\end{equation}
In the following section, our goal is to understand the effect of the environment on the quantum battery parameters by considering the interaction between the battery and the massless quantum scalar field while the charging process is going on.
\section{Derivation of the Master Equation}\label{sec:GKSL}
\noindent In this section, we derive the \textit{Gorini-Kossakowski-Sudarshan-Lindblad (GKSL)} quantum master equation \cite{GKS_1976, Lindblad1976} for the Unruh-DeWitt battery, considering the quadratic coupling interactions. The GKSL master equation has been previously studied in the context of a single \cite{Benatti_2004, yu2008understanding, hu2012geometric, jin2014dynamical,liu2021relativistic} and two Unruh-DeWitt detectors \cite{Benatti_2004,zhang2007unruh,hu2015entanglement,yang2016entanglement,cheng2018entanglement,she2019entanglement,zhou2021entanglement, chen2022entanglement}, primarily assuming a linear coupling between the detectors and the quantum field. The model based on linear coupling was initially introduced in this context in \cite{DeWitt1979, Unruh1984, birrell1984quantum}, and also recently studied in different aspects to yield certain fascinating results \cite{Chatterjee2021_1, Mukherjee2022, Mukherjee2023, Wu2023}.

In this study, we consider the open quantum framework as the system (UDW battery) always interacts with the environment (background quantum field). Here UDW battery shows the evolution in the proper time $\tau$. During this time, battery is travelling along a spatial trajectory in Minkowski spacetime and absorbs charging from the external classical pulse. In the laboratory frame, we assume that the trajectory of the moving battery is given by a worldline function $x(\tau)=(t(\tau),\mathbf{x}(\tau))$.

To streamline the derivation, we first move to the interaction picture, which allows us to decouple the free evolution and concentrate solely on the effects of the interaction. Hence, the complete Hamiltonian of the detector-field system in the interaction picture takes the form
\begin{equation}
\widetilde{H}(\tau)=\widetilde{H}_{B}(\tau)+\widetilde{H}_{F}(\tau)+\widetilde{H}_{I}(\tau)
\end{equation}
where $\widetilde{H}_{B}(\tau)$ is the Hamiltonian of the battery, $\widetilde{H}_{F}(\tau)$ is the Hamiltonian of the fluctuating scalar field $\phi(x(\tau))$ in Minkowski spacetime, and $\widetilde{H}_{I}(\tau)$ is the interaction Hamiltonian between the battery and the quantum scalar field.

In the interaction picture, the battery Hamiltonian and the field Hamiltonian become
\begin{eqnarray}
    \widetilde{H}_{B}(\tau)&=&H_{B}=\frac{\alpha}{2}(\sigma_{+} +\sigma_{-})\\
    \widetilde{H}_{F}(\tau)&=&H_{F}=\int d^{3}\mathbf{k}\,\omega_{\mathbf{k}}\,a_{\mathbf{k}}^{\dagger}a_{\mathbf{k}}~.
\end{eqnarray}
Following Ref. \cite{Hummer2016, Sachs2017, Gray2018, Wu2023}, initially we consider the interaction Hamiltonian between the battery and the quantum scalar field as
\begin{equation}\label{quad_Ham0}
    \widetilde{H}_{I}(\tau)=\mu\big(\widetilde{\sigma}_{+}(\tau)+\widetilde{\sigma}_{-}(\tau)\big):\phi^2[x(\tau)]:
\end{equation}
where $\mu<<1$ and $\phi^{2}[x(\tau)]$ is normal-ordered.
Now using the mode expansion of the massless scalar field $\phi[x(\tau)]$
\begin{equation}\label{phi}
\phi[x(\tau)]=\frac{1}{(2\pi)^{3/2}}\int \frac{d^3\mathbf{k}}{\sqrt{ 2|\mathbf{k}|}}\left(\hat{a}_{\mathbf{k}} e^{-ik\cdot x(\tau)}+\text{H.c}\right)
\end{equation}
and after taking normal ordering, $:\phi^{2}[(x(\tau)]:$ takes the form
\begin{equation}\label{phi2}
    :\phi^{2}[(x(\tau)]: \,= \frac{1}{(2\pi)^3}\int \frac{d^3\mathbf{k}\,d^3\mathbf{k'}}{\sqrt{ 4|\mathbf{k}||\mathbf{k'}|}} \left( \hat{a}_{\mathbf{k}}\hat{a}_{\mathbf{k'}} e^{-i(k+k')\cdot x(\tau)}+2\hat{a}^{\dagger}_{\mathbf{k}}\hat{a}_{\mathbf{k'}} e^{i(k-k')\cdot x(\tau)}+\hat{a}^{\dagger}_{\mathbf{k}}\hat{a}^{\dagger}_{\mathbf{k'}} e^{i(k+k')\cdot x(\tau)}\right)
\end{equation}
where $k\cdot x(\tau)=|\mathbf{k}| t(\tau) - \mathbf{k} \cdot \mathbf{x}(\tau)$.
The normal ordering carried out in eq.~\eqref{phi2} is done in order to get rid of divergences in the detector’s vacuum excitation probability \cite{hinton1984particle} due to the quadratic field operator. It was shown in \cite{Hummer2016, Sachs2017} that if the non-linear interaction is normal ordered, then no such divergences appear. This ensures that the Born approximation can be safely carried out in the following derivation.

Following the interaction Hamiltonian given (in eq. ($3$)) in Ref. \cite{gilles1993tp}, and using the similar rotating wave approximation, the interaction Hamiltonian in the interaction picture (eq.~\eqref{quad_Ham0}) turns out to be
\begin{equation}\label{quad_Ham}
    \widetilde{H}_{I}(\tau)= \frac{\mu}{(2\pi)^3}\int \frac{d^3\mathbf{k}\,d^3\mathbf{k'}}{\sqrt{ 4|\mathbf{k}||\mathbf{k'}|}}\left(\widetilde{\sigma}_{+}(\tau)\,\hat{a}_{\mathbf{k}}\hat{a}_{\mathbf{k'}} e^{-i(k+k')\cdot x(\tau)}+\widetilde{\sigma}_{-}(\tau)\,\hat{a}^{\dagger}_{\mathbf{k}}\hat{a}^{\dagger}_{\mathbf{k'}} e^{i(k+k')\cdot x(\tau)}\right)~.
\end{equation}
Here 
\begin{align}\label{def_sigma}
\widetilde{\sigma}_{\pm}(\tau)&=e^{iH_{B}{\tau}}\sigma_{\pm}e^{-iH_{B}{\tau}}=s\mp p\,e^{i\alpha\tau}\pm q\,e^{-i\alpha\tau}
\end{align}
with $s=\sigma_1/2, \,p=(\sigma_3-i\sigma_2)/4,\, q=(\sigma_3+i\sigma_2)/4$.\\[2pt]
Note that the rotating wave approximation used in writing eq.~\eqref{quad_Ham} corresponds to keeping the energy conserving terms in the interaction Hamiltonian given by eq.~\eqref{quad_Ham0} \cite{carmichael1999statistical, scully1997quantum}. Clearly, the first term in eq.~\eqref{quad_Ham} corresponds to the process in which an atom makes a transition from the lower level to the upper level with the annihilation of two quanta of the field. Similarly, the second term in eq.~\eqref{quad_Ham} corresponds to the process in which an atom makes a transition from the upper level to the lower level with the creation of two quanta of the field. Note that the rotating wave approximation can be also thought of keeping terms in the interaction Hamiltonian that oscillates with frequency $(\omega_0-\omega)$, where $\omega_0$ is the frequency of the atom and $\omega$ is the frequency of the field. Both the terms in eq.~\eqref{quad_Ham} are of this form since $(k^{0}+k^{'0})$ arising from the $(k+k')$ is the frequency of the field and the atomic frequency arising from the $\widetilde{\sigma}_{+}(\tau)$ and $\widetilde{\sigma}_{-}(\tau)$ come with the opposite sign of $(k^{0}+k^{'0})$.

Hence, the equation of motion of the total system is given by 
\begin{equation}\label{Von1}
    \frac{\partial\widetilde{\rho}_{tot}(\tau)}{\partial\tau}=-i\Big[\widetilde{H}_{I}(\tau),\widetilde{\rho}_{tot}(\tau)\Big]~.
\end{equation}
The solution of the eq.~\eqref{Von1} can be written as
\begin{eqnarray}\label{rho_sol1}
    \widetilde{\rho}_{tot}(\tau)=\widetilde{\rho}_{tot}(0)-i\int_{0}^{\tau}du \Big[\widetilde{H}_{I}(u),\widetilde{\rho}_{tot}(u)\Big]~.
\end{eqnarray}
Putting eq.~\eqref{rho_sol1} once again in the eq.~\eqref{Von1} and taking the partial trace over field degrees of freedom, up to $\mathcal{O}(\mu^2)$ we get
\begin{eqnarray}\label{q_main_eq}
    \frac{\partial\widetilde{\rho}_{s}(\tau)}{\partial\tau}=-i\,Tr_{F}\big[\widetilde{H}_{I}(\tau),\widetilde{\rho}_{tot}(0)\big]-\int_{0}^{\tau}du\,\,Tr_{F}\Big[\widetilde{H}_{I}(\tau),\big[\widetilde{H}_{I}(u),\widetilde{\rho}_{tot}(u)\big]\Big]~.
\end{eqnarray}
Let us now assume that at the initial time, the total density matrix of the detector-field system is given by \(\widetilde{\rho}_{\text{tot}}(0) = \widetilde{\rho}_s(0) \otimes |0\rangle\langle 0|\), where \(\rho_s^{I}(0)\) represents the initial reduced density matrix of the detector, and \(|0\rangle\langle 0|\) denotes the vacuum state of the field.
Using this in the first term of the R.H.S of the eq.~\eqref{q_main_eq} and calculating the first term we get $Tr_{F}\big[\widetilde{H}_{I}(\tau),\widetilde{\rho}_{tot}(0)\big]=0$.
Now invoking the \textit{Born approximation} \cite{Breuer2007}, which asserts that when the bath is significantly larger than the system, its influence on the system's dynamics can be neglected, it can be written, 
\begin{eqnarray}
    \widetilde{\rho}_{tot}(u)\approx \widetilde{\rho}_{s}(u)\otimes\rho_{F}~.
\end{eqnarray}
Changing the variable $u\rightarrow\tau'$ such that $\tau'=\tau-u$ we get $\int_{0}^{\tau}du=\int_{0}^{\tau}d\tau'$. Therefore, eq.~\eqref{q_main_eq} becomes
\begin{equation}\label{q_main_eq1}
    \frac{\partial\widetilde{\rho}_{s}(\tau)}{\partial\tau}
    =-\int_{0}^{\tau}d{\tau'}\,Tr_{F}\Big[\widetilde{H}_{I}(\tau),\big[\widetilde{H}_{I}(\tau-\tau'),\widetilde{\rho}_{s}(\tau-\tau')\otimes\rho_{F}\big]\Big]~.
\end{equation}
Using the form of the interaction Hamiltonian (eq.~\eqref{quad_Ham}) in the eq.~\eqref{q_main_eq1}, expanding the double commutator carefully, and computing the each term straightforwardly, we can recast eq.~\eqref{q_main_eq1} as
\begin{align}\label{q_main_eq2}
    \frac{\partial\widetilde{\rho}_{s}(\tau)}{\partial\tau}
=&\mu^2\int_{0}^{\tau}d{\tau'}\bigg[\Big(\widetilde{\sigma}_{-}(\tau-\tau')\widetilde{\rho}_{s}(\tau-\tau')\widetilde{\sigma}_{+}(\tau)-\widetilde{\sigma}_{+}(\tau)\widetilde{\sigma}_{-}(\tau-\tau')\widetilde{\rho}_{s}(\tau-\tau')\Big)G^{+(2)}(\tau,\tau-\tau')\nonumber\\
    +&\Big(\widetilde{\sigma}_{-}(\tau)\widetilde{\rho}_{s}(\tau-\tau')\widetilde{\sigma}_{+}(\tau-\tau')-\widetilde{\rho}_{s}(\tau-\tau')\widetilde{\sigma}_{+}(\tau-\tau')\widetilde{\sigma}_{-}(\tau)\Big)\big(G^{+(2)}(\tau,\tau-\tau')\big)^{*}\bigg]~.
\end{align}
Here, we have defined
\begin{eqnarray}
    G^{+(2)}(\tau,\tau-\tau')&\equiv&\langle 0\vert:\phi^{2}(x(\tau)):\,:\phi^{2}(x(\tau-\tau')):\vert 0\rangle~\label{defg2}\\
    \big(G^{+(2)}(\tau,\tau-\tau')\big)^{*}&\equiv& \langle 0\vert:\phi^{2}(x(\tau-\tau')):\,:\phi^{2}(x(\tau)):\vert 0\rangle~.~\label{defg2s}
\end{eqnarray}
The superscript $(2)$ indicates the quadratic coupling Wightman function, to distinguish it from the usual linear coupling Wightman function $G^{+}(\tau,\tau-\tau')$.\\
Now assuming that the bath correlation time \(\tau_F\) is much shorter than the characteristic system timescale \(\tau\) (i.e.\(\tau \gg \tau_F\)), the system’s density matrix \(\widetilde{\rho}_s(\tau - \tau')\) can be replaced by \(\widetilde{\rho}_s(\tau)\). This allows us to extend the upper limit of the integral to infinity without altering its value. This is known as the \textit{Markov approximation} \cite{Breuer2007}. At this point we would like to mention that even for the non-linear interaction, \textit{the Born-Markov approximation} can be carried out safely. This was pointed out in  \cite{gilles1993tp}.

Therefore, using the form of $\widetilde{\sigma}_{\pm}(\tau)$ given in eq.~\eqref{def_sigma} in the eq.~\eqref{q_main_eq2} under this Markovian approximation, we can recast eq.~\eqref{q_main_eq2} as
\begin{align}\label{q_eq_final1}
   \frac{\partial\widetilde{\rho}_{s}(\tau)}{\partial\tau}&=\mu^2\Big[\big(q\widetilde{\rho}_{s}(\tau)p-pq\widetilde{\rho}_{s}(\tau)\big)\,\mathcal{J}_1+\big(p\widetilde{\rho}_{s}(\tau)q-qp\widetilde{\rho}_{s}(\tau)\big)\,\mathcal{J}_2+\big(p\widetilde{\rho}_{s}(\tau)q-\widetilde{\rho}_{s}(\tau)qp\big)\,\mathcal{J}_3\nonumber\\
   &+\big(q\widetilde{\rho}_{s}(\tau)p-\widetilde{\rho}_{s}(\tau)pq\big)\,\mathcal{J}_4+\big(s\widetilde{\rho}_{s}(\tau)s-s^2\widetilde{\rho}_{s}(\tau)\big)\,\mathcal{J}_5+\big(s\widetilde{\rho}_{s}(\tau)s-\widetilde{\rho}_{s}(\tau)s^2\big)\,\mathcal{J}_6\Big]
\end{align}
with
\begin{eqnarray}\label{q_int_main}
\mathcal{J}_1&=&\int_{0}^{\infty}d{\tau'}e^{i \alpha \tau'}\,G^{+(2)}(\tau,\tau-\tau')~,\qquad \mathcal{J}_2=\int_{0}^{\infty}d{\tau'}e^{-i \alpha \tau'}\,G^{+(2)}(\tau,\tau-\tau')~,\nonumber\\
\mathcal{J}_3&=&\int_{0}^{\infty}d{\tau'}e^{i \alpha \tau'}\,\big(G^{+(2)}(\tau,\tau-\tau')\big)^{*}~,\qquad 
\mathcal{J}_4=\int_{0}^{\infty}d{\tau'}e^{-i \alpha \tau'}\,\big(G^{+(2)}(\tau,\tau-\tau')\big)^{*}~,\nonumber\\
\mathcal{J}_5&=&\int_{0}^{\infty}d{\tau'}\,G^{+(2)}(\tau,\tau-\tau')~,\qquad
\mathcal{J}_6=\int_{0}^{\infty}d{\tau'}\,\big(G^{+(2)}(\tau,\tau-\tau')\big)^{*}~.
\end{eqnarray}

Going to the Fourier space, eq.~\eqref{q_int_main} we get
\begin{eqnarray}\label{q_int_result}
\mathcal{J}_1&=&\frac{1}{2}\Big[\mathcal{G}^{(2)}(\alpha)+\mathcal{K}^{(2)}(\alpha)\Big]~,\qquad \mathcal{J}_2=\frac{1}{2}\Big[\mathcal{G}^{(2)}(-\alpha)+\mathcal{K}^{(2)}(-\alpha)\Big]~,\nonumber\\
\mathcal{J}_3&=&\frac{1}{2}\Big[\mathcal{G}^{(2)}(-\alpha)-\mathcal{K}^{(2)}(-\alpha)\Big]~,\qquad
\mathcal{J}_4=\frac{1}{2}\Big[\mathcal{G}^{(2)}(\alpha)-\mathcal{K}^{(2)}(\alpha)\Big]~,\nonumber\\
\mathcal{J}_5&=&\frac{1}{2}\Big[\mathcal{G}^{(2)}(0)+\mathcal{K}^{(2)}(0)\Big]~,\qquad
\mathcal{J}_6=\frac{1}{2}\Big[\mathcal{G}^{(2)}(0)-\mathcal{K}^{(2)}(0)\Big]~,
\end{eqnarray}
where 
\begin{eqnarray}
    \mathcal{G}^{(2)}(\lambda)&=&\int_{-\infty}^{\infty}d{\tau'}e^{i \lambda \tau'}\,G^{+(2)}(\tau,\tau-\tau')\\
    \mathcal{K}^{(2)}(\lambda)&=&\frac{\mathcal{P}}{\pi i}\int_{-\infty}^{\infty}d{\omega}\frac{\mathcal{G}^{(2)}(\omega)}{\omega-\lambda}~.
\end{eqnarray}
The value of $\mathcal{G}^{(2)}(\lambda)$ will be calculated later.
Putting eq.~\eqref{q_int_result} into eq.~\eqref{q_eq_final1} and rearranging, we obtain
\begin{align}\label{q_eq_final2}
    \frac{\partial\widetilde{\rho}_{s}(\tau)}{\partial\tau}
    &=\frac{\mu^2}{2}\Big[\mathcal{G}^{(2)}(\alpha)\big(2q\widetilde{\rho}_{s}(\tau)p-pq\widetilde{\rho}_{s}(\tau)-\widetilde{\rho}_{s}(\tau)pq\big)+\mathcal{G}^{(2)}(-\alpha)\big(2p\widetilde{\rho}_{s}(\tau)q-qp\widetilde{\rho}_{s}(\tau)-\widetilde{\rho}_{s}(\tau)qp\big)\nonumber\\
    \,\,\,&+\mathcal{G}^{(2)}(0)\big(2s\widetilde{\rho}_{s}(\tau)s-s^{2}\widetilde{\rho}_{s}(\tau)-\widetilde{\rho}_{s}(\tau)s^{2}\big)+\mathcal{K}^{(2)}(\alpha)\big(\widetilde{\rho}_{s}(\tau)pq-pq\widetilde{\rho}_{s}(\tau)\big)\nonumber\\
    &\,\,+\mathcal{K}^{(2)}(-\alpha)\big(\widetilde{\rho}_{s}(\tau)qp-qp\widetilde{\rho}_{s}(\tau)\big)+\mathcal{K}^{(2)}(0)\big(\widetilde{\rho}_{s}(\tau)s^{2}-s^{2}\widetilde{\rho}_{s}(\tau)\big)\Big]~.
\end{align}
Here $\mathcal{K}^{(2)}(0)\big(\widetilde{\rho}_{s}(\tau)s^{2}-s^{2}\widetilde{\rho}_{s}(\tau)\big)=0$ as $s^{2}\sim \mathds{I}$.

\noindent In the Schr\"{o}dinger picture, above eq.~\eqref{q_eq_final2} turns out to be
\begin{align}\label{q_eq_final3}
    &\,\,\frac{\partial\rho_{s}(\tau)}{\partial\tau}+i\big[H_B,\rho_{s}(\tau)\big]\nonumber\\
    &=\frac{\mu^2}{2}\Big[\mathcal{G}^{(2)}(\alpha)\big(2q\rho_{s}(\tau)p-pq\rho_{s}(\tau)-\rho_{s}(\tau)pq\big)+\mathcal{G}^{(2)}(-\alpha)\big(2p\rho_{s}(\tau)q-qp\rho_{s}(\tau)-\rho_{s}(\tau)qp\big)\nonumber\\
    &+\mathcal{G}^{(2)}(0)\big(2s\rho_{s}(\tau)s-s^{2}\rho_{s}(\tau)-\rho_{s}(\tau)s^{2}\big)+\mathcal{K}^{(2)}(\alpha)\big(\rho_{s}(\tau)pq-pq\rho_{s}(\tau)\big)\nonumber\\
    &+\mathcal{K}^{(2)}(-\alpha)\big(\rho_{s}(\tau)qp-qp\rho_{s}(\tau)\big)\Big]~.
\end{align}
In terms of Pauli matrices, eq.~\eqref{q_eq_final3} can be rewritten in the form
\begin{equation}\label{q_GKSL}
\frac{\partial \rho_s(\tau)}{\partial \tau}=-i\Big[\mathcal{H}_{eff}, \rho_s(\tau)\Big] + \mathcal{D}\big[\rho_s(\tau)\big]~.
\end{equation}
This represents the \textit{GKSL master equation} governing the evolution of the reduced density matrix $\rho_{s}(\tau)$ for the Unruh-DeWitt battery, quadratically coupled to a massless quantum scalar field. The effective Hamiltonian $\mathcal{H}_{eff}$ and the dissipator $\mathcal{D}\big[\rho_s(\tau)\big]$ are given by
\begin{eqnarray}\label{q_H_eff,Lind}
\mathcal{H}_{eff}&=&\frac{1}{2}\left[\alpha+\underbrace{\frac{i\mu^2}{8}\Big\{\mathcal{K}^{(2)}(-\alpha)-\mathcal{K}^{(2)}(\alpha)\Big\}}_{\Omega_{LS}}\right](\sigma_{+}+\sigma_{-})\nonumber\\
&=&\frac{1}{2}\Big[\alpha+\Omega_{LS}\Big](\sigma_{+}+\sigma_{-})\\
\mathcal{D}\big[\rho_s(\tau)\big]&=&\frac{1}{2}\displaystyle\sum_{i,j=1}^{3} \Gamma_{ij}\Big[2\sigma_{j}\rho_s(\tau)\sigma_{i}-\sigma_i\sigma_j\rho_s(\tau)-\rho_s(\tau)\sigma_i\sigma_j\Big]~.\nonumber\\
\end{eqnarray}
The structure of the \textit{GKSL master equation} for the quadratic coupling (eq.~\eqref{q_GKSL}) closely mirrors that of the linear coupling case (eq.~\eqref{l_GKSL}), differing primarily in the form of the effective Hamiltonian and the Kossakowski matrix coefficients $\Gamma_{ij}$. In the quadratic coupling case, Kossakowski matrix coefficients $\Gamma_{ij}$ is given by
\begin{equation}
\Gamma_{ij}=\mathcal{R}\delta_{ij}-i\mathcal{S}\varepsilon_{ijk}\delta_{k1}-\mathcal{T}\delta_{i1}\delta_{j1}~
\end{equation}
where 
\begin{eqnarray}\label{R,S,T}
    \mathcal{R}&=&\frac{\mu^2}{16}\Big[\mathcal{G}^{(2)}(\alpha)+\mathcal{G}^{(2)}(-\alpha)\Big]\,\nonumber\\
    \mathcal{S}&=&\frac{\mu^2}{16}\Big[\mathcal{G}^{(2)}(\alpha)-\mathcal{G}^{(2)}(-\alpha)\Big]\,\nonumber\\
    \mathcal{T}&=&\frac{\mu^2}{4}\mathcal{G}^{(2)}(0)-\mathcal{R}~.
\end{eqnarray}
Here, we consider the Lamb shift correction term $\Omega_{LS}<<\alpha$, therefore we neglect the Lamb shift correction term in the following calculations.
\section{Effect of the environment on the  Unruh-DeWitt battery: Quadratic Coupling Scenario}\label{sec:Env_effect}
\noindent We consider the situation where the battery  is moving along a nontrivial trajectory  with a constant square of magnitudes of four acceleration $a_\mu a^\mu =a^2$ \cite{Abdolrahimi2014, Liu2021, Hao2023}. In the case where the battery follows a trajectory composed of uniform acceleration along one direction, combined with constant four-velocity components in the orthogonal plane to the acceleration, one can write the worldline function as
\begin{equation}\label{trajectory}
x(\tau)=\left(\frac{a}{\xi^2}\sinh(\xi\tau),\,\frac{a}{\xi^2}\cosh(\xi\tau),\,w_1 \tau,\,w_2\tau\right)
\end{equation}
where $w_1$, $w_2$ are the constant components of the four velocity, and the parameter $\xi=\frac{a}{\sqrt{1+\mathcal{u}^2}}$, where $\mathcal{u}^2=w_1^2+w_2^2$. This kind of trajectory can be seen in a physical scenario, if we place a charged atom (with $\frac{e}{m}=1$) in a frame where a uniform electric field acting in the $x$ direction, and the initial velocity of the atom is in $y$ and $z$ directions \cite{Abdolrahimi2014}. 

\noindent Building on the insights from Refs. \cite{Hummer2016, Sachs2017, Gray2018, Wu2023}, this study primarily focuses on the scenario where the quantum scalar field interacts with the detector via a quadratic coupling. 

\noindent The interaction Hamiltonian in the interaction picture is given by eq.~\eqref{quad_Ham}
\begin{equation}
    \widetilde{H}_{I}(\tau)= \frac{\mu}{(2\pi)^3}\int \frac{d^3\mathbf{k}\,d^3\mathbf{k'}}{\sqrt{ 4|\mathbf{k}||\mathbf{k'}|}}\left(\widetilde{\sigma}_{+}(\tau)\,\hat{a}_{\mathbf{k}}\hat{a}_{\mathbf{k'}} e^{-i(k+k')\cdot x(\tau)}+\widetilde{\sigma}_{-}(\tau)\,\hat{a}^{\dagger}_{\mathbf{k}}\hat{a}^{\dagger}_{\mathbf{k'}} e^{i(k+k')\cdot x(\tau)}\right)~.
\end{equation}
For the quadratic coupling, using the Wick's theorem the Wightman function takes the form \cite{Sachs2017}
\begin{align}
&\,\,G^{+(2)}(\tau,\tau-\tau')=\langle 0\vert :\phi^{2}(x(\tau))::\phi^{2}(x(\tau-\tau')):\vert 0\rangle\nonumber\\
&=2\Big(\langle 0\vert\phi(x(\tau))\phi(x(\tau-\tau'))\vert 0\rangle\Big)^2~.
\end{align}
Using the mode expansion of the massless scalar field given in eq.~\eqref{phi}, the quadratic field Wightman function $G^{+(2)}(\tau,\tau-\tau')$ becomes
\begin{align}
&G^{+(2)}(\tau,\tau-\tau')=\langle 0\vert :\phi^{2}(x(\tau))::\phi^{2}(x(\tau-\tau')):\vert 0\rangle\nonumber\\
&=\frac{1}{8\pi^4}\frac{1}{\big[\big(t(\tau)-t(\tau-\tau')-i\epsilon\big)^2-\vert \textbf{x}(\tau)-\textbf{x}(\tau-\tau')\vert^2\big]^2}~.
\end{align}
Using the trajectory given in eq.~\eqref{trajectory}, the quadratic field Wightman function along this trajectory turns out to be
\begin{equation}\label{Wightman1A}
G^{+(2)}(\tau,\tau-\tau')=\frac{a^4}{128\pi^4 (1+\mathcal{u}^2)^4}\left[\sinh^2\left(\frac{a(\tau')}{2\sqrt{1+\mathcal{u}^2}}-i\epsilon\right)-\frac{\mathcal{u}^2a^2(\tau')^2}{4(1+\mathcal{u}^2)^2}\right]^{-2}~.
\end{equation} 
As in the earlier case, in order to gain physical insights we consider the limits: (i) $\mathcal{u}^2<<1$ (non-relativistic limit) and (ii) $\mathcal{u}^2>>1$ (relativistic limit) for calculating the Fourier transform of the field correlation function  $\mathcal{G}^{(2)}(\pm \alpha)$.

\noindent In the limit $\mathcal{u}^2<<1$, the Wightman function upto $\mathcal{O}(\mathcal{u}^2)$ takes the form

\begin{align}\label{Wightman1A_NR}
G^{+(2)}(\tau,\tau-\tau')&=\frac{a^4}{128\pi^4}\left[\frac{(1-4\mathcal{u}^2)}{\sinh^4\left(\frac{a(\tau')}{2}-i\epsilon\right)}+\frac{a^2(\tau')^2+a(\tau')\sinh\big(a(\tau')-i\epsilon\big)}{2\sinh^6\left(\frac{a(\tau')}{2}-i\epsilon\right)}\mathcal{u}^2\right]~.
\end{align}

\noindent Now calculating the Fourier transform $\mathcal{G}^{(2)}(\pm \alpha)$ by using the above Wightman function eq.~\eqref{Wightman1A_NR}, we get
\begin{align}\label{g2a1}
    \mathcal{G}^{(2)}(\alpha)&=\frac{\alpha^3}{24\pi^3}\left(1+\frac{a^2}{\alpha^2}\right)\frac{e^{2\pi\alpha/a}}{(e^{2\pi\alpha/a}-1)}+\frac{a^3 \mathcal{u}^{2} e^{2\pi\alpha/a}}{120\pi^2(e^{2\pi\alpha/a}-1)^2}\left[8+\frac{35\alpha^2}{a^2}+\frac{15\alpha^4}{a^4}\right.\nonumber\\
&\left.-\frac{2\pi\alpha}{a}\left(4+\frac{5\alpha^2}{a^2}+\frac{\alpha^4}{a^4}\right)\coth\left(\frac{\pi\alpha}{a}\right)\right]~.
\end{align}
\begin{align}\label{g2a2}
    \mathcal{G}^{(2)}(-\alpha)&=\frac{\alpha^3}{24\pi^3}\left(1+\frac{a^2}{\alpha^2}\right)\frac{1}{(e^{2\pi\alpha/a}-1)}+\frac{a^3 \mathcal{u}^{2} e^{2\pi\alpha/a}}{120\pi^2(e^{2\pi\alpha/a}-1)^2}\left[8+\frac{35\alpha^2}{a^2}+\frac{15\alpha^4}{a^4}\right.\nonumber\\
&\left.-\frac{2\pi\alpha}{a}\left(4+\frac{5\alpha^2}{a^2}+\frac{\alpha^4}{a^4}\right)\coth\left(\frac{\pi\alpha}{a}\right)\right]~.
\end{align}
In the limit $\alpha\rightarrow0$, $\mathcal{G}^{(2)}(\alpha)$ takes the form
\begin{equation}\label{g201}
    \mathcal{G}^{(2)}(0)=\frac{a^3}{48\pi^4}\left[1-\left(\frac{5}{2}-\frac{4\pi^2}{15}\right)\mathcal{u}^2\right]~.
\end{equation}
Substituting eq.(s) (\ref{g2a1}, \ref{g2a2}, \ref{g201}) into eq.~\eqref{R,S,T}, we get
\begin{eqnarray}\label{ABC1A_Nonrelativistic}
\mathcal{R}&=&\frac{\mu^2\alpha^3}{384\pi^3}\left(1+\frac{a^2}{\alpha^2}\right)\left[\coth\left(\frac{\pi\alpha}{a}\right)-\frac{2\pi}{\alpha^3} \mathcal{j}(\alpha, a)\mathcal{u}^2\right]\nonumber\\ \mathcal{S}&=&\frac{\mu^2\alpha^3}{384\pi^3}\left(1+\frac{a^2}{\alpha^2}\right)\nonumber\\
\mathcal{T}&=&\frac{\mu^2a^3}{192\pi^4}\left[1-\left(\frac{5}{2}-\frac{4\pi^2}{15}\right)\mathcal{u}^2\right]-\mathcal{R}
\end{eqnarray}
where 
\begin{align}
\mathcal{j}(\alpha, a)=&\frac{a^3 e^{2\pi\alpha/a}}{5\left(1+\frac{a^2}{\alpha^2}\right)(e^{2\pi\alpha/a}-1)^2}\left[8+\frac{35\alpha^2}{a^2}+\frac{15\alpha^4}{a^4}-\frac{2\pi\alpha}{a}\left(4+\frac{5\alpha^2}{a^2}+\frac{\alpha^4}{a^4}\right)\coth\left(\frac{\pi\alpha}{a}\right)\right]~.
\end{align}

In the limit $\mathcal{u}^2>>1$, the Wightman function can be approximated upto $\mathcal{O}(\frac{1}{\mathcal{u}^8})$ as
\begin{equation}\label{Wightman1A_UR}
G^{+(2)}(\tau,\tau-\tau')=\frac{a^4}{128\pi^4}\left[\frac{1}{\sinh^4\left(\frac{a(\tau')}{2\mathcal{u}}-i\epsilon\right)\mathcal{u}^8}\right]\,.
\end{equation}
Calculating $\mathcal{G}^{(2)}(\pm \alpha)$ by using the above Wightman function eq.~\eqref{Wightman1A_UR}, we get
\begin{equation}\label{g2a3}
    \mathcal{G}^{(2)}(\alpha)=\frac{\alpha^3}{24\pi^3\mathcal{u}^4}\left(1+\frac{a^2}{\alpha^2\mathcal{u}^2}\right)\frac{e^{2\pi\alpha\mathcal{u}/a}}{(e^{2\pi\alpha\mathcal{u}/a}-1)}
\end{equation}
\begin{equation}\label{g2a4}
    \mathcal{G}^{(2)}(-\alpha)=\frac{\alpha^3}{24\pi^3\mathcal{u}^4}\left(1+\frac{a^2}{\alpha^2\mathcal{u}^2}\right)\frac{1}{(e^{2\pi\alpha\mathcal{u}/a}-1)}
\end{equation}
with, \begin{equation}\label{g202}
\mathcal{G}^{(2)}(0)=\frac{\alpha^3}{48\pi^4\mathcal{u}^7}~.
\end{equation}
Putting eq.(s) (\ref{g2a3}, \ref{g2a4}, \ref{g202}) into eq.~\eqref{R,S,T}, we get
\begin{align}\label{ABC1A_Ultrarelativistic}
\mathcal{R}&=\frac{\mu^2\alpha^3}{384\pi^3\mathcal{u}^4}\left(1+\frac{a^2}{\alpha^2\mathcal{u}^2}\right)\coth\left(\frac{\pi\alpha\mathcal{u}}{a}\right)\nonumber\\
\mathcal{S}&=\frac{\mu^2\alpha^3}{384\pi^3\mathcal{u}^4}\left(1+\frac{a^2}{\alpha^2\mathcal{u}^2}\right)\nonumber\\
\mathcal{T}&=\frac{\mu^2a^3}{192\pi^4\mathcal{u}^7}-\mathcal{R}
\end{align}
In the following section, we are going to study the relativistic effects on the performance of the battery while it is moving along the trajectory given by eq.~\eqref{trajectory}.
\section{Dynamics of the Unruh-DeWitt battery}\label{sec:Dyn_effect}
 \noindent Our approach here is to study the dynamics of the quantum battery by solving the GKSL master equation for linear coupling (eq.~\eqref{l_GKSL}) and for quadratic coupling (eq.~\eqref{q_GKSL}) scenario. The underlying motivation of this analysis serves as a foundation for a future comparative study, intended to reveal which interaction model more effectively enhances the battery's performance. 
 
 For doing so, at first we express the evolved state of the battery in terms of the components of the Bloch vector $\vec{r}(\tau)$ as 
 \begin{eqnarray}\label{rho_s}
 \rho_{s}(\tau)=\frac{1}{2}\Big(\mathds{I}+\displaystyle\sum_{i=1}^3 r_{i}(\tau)\sigma_i\Big)~. 
 \end{eqnarray}
 Substituting the above form of $\rho_{s}(\tau)$ (eq.~\eqref{rho_s}) in eq.~\eqref{l_GKSL} and eq.~\eqref{q_GKSL}, and solving it, we get the time evolved Bloch vector for the two cases in the form
\begin{align}\label{l_Bloch}
\vec{r}_{lin}(\tau)=&\left(\frac{\mathcal{B}}{\mathcal{A}}(e^{-4\mathcal{A}\tau}-1),\,e^{-2(2\mathcal{A}+\mathcal{C})\tau}\sin(\alpha\tau),-e^{-2(2\mathcal{A}+\mathcal{C})\tau}\cos(\alpha\tau)\right)
\end{align}
and 
\begin{align}\label{q_Bloch}
\vec{r}_{quad}(\tau)=&\left(\frac{\mathcal{S}}{\mathcal{R}}(e^{-4\mathcal{R}\tau}-1),\,e^{-2(2\mathcal{R}+\mathcal{T})\tau}\sin(\alpha\tau),-e^{-2(2\mathcal{R}+\mathcal{T})\tau}\cos(\alpha\tau)\right)
\end{align}
 with
\begin{equation}\label{l_Bloch_Norm}
r_{lin}\equiv \vert \vec{r}_{lin}(\tau)\vert=\left[\frac{\mathcal{B}^2(e^{-4\mathcal{A}\tau}-1)^2}{\mathcal{A}^2} +e^{-4(2\mathcal{A}+\mathcal{C})\tau}\right]^{1/2}\,
\end{equation}
and 
\begin{equation}\label{q_Bloch_Norm}
r_{quad}\equiv \vert \vec{r}_{quad}(\tau)\vert=\left[\frac{\mathcal{S}^2(e^{-4\mathcal{R}\tau}-1)^2}{\mathcal{R}^2} +e^{-4(2\mathcal{R}+\mathcal{T})\tau}\right]^{1/2}\,.
\end{equation}
For both the cases, we have taken the initial state of the battery as the ground state $\vert g\rangle$, and the initial value of the Bloch vector becomes $\vec{r}_{lin}(0)=\vec{r}_{quad}(0)=(0,\,0,\,-1)$. Using this initial value of the Bloch vector, we get the final results given in eq.(s)~(\ref{l_Bloch}, \ref{q_Bloch}).

To obtain the important parameters which are associated with the quantum battery performance, we need to find the optimal unitary operations $U_\sigma$ and $U_\pi$ which will transform the battery state $\rho_{s}(\tau)$ into a passive state $\sigma_{s}(\tau)$ and an active state $\pi_{s}(\tau)$, respectively.

Considering the following form of $U_\sigma$ and $U_\pi$ given by
\begin{equation}
U_\sigma=
\begin{bmatrix}
\frac{r_1+i r_2}{\sqrt{2r(r+r_3)}}&  -\sqrt{\frac{r+r_3}{2r}} \\
\sqrt{\frac{r+r_3}{2r}}&      \frac{r_1-i r_2}{\sqrt{2r(r+r_3)}}
\end{bmatrix}
~,\qquad
U_\pi=
\begin{bmatrix}
\sqrt{\frac{r+r_3}{2r}}&      \frac{r_1-i r_2}{\sqrt{2r(r+r_3)}}\\
\frac{r_1+i r_2}{\sqrt{2r(r+r_3)}}&  -\sqrt{\frac{r+r_3}{2r}}
\end{bmatrix}
,
\end{equation}
the passive and active state corresponding to the battery state $\rho_{s}(\tau)$ takes the form
\begin{equation}
\sigma_{s}(\tau)=
\begin{pmatrix}
\frac{1-r}{2}& 0\\
0& \frac{1+r}{2}
\end{pmatrix}
~,\qquad
\pi_{s}(\tau)=
\begin{pmatrix}
\frac{1+r}{2}& 0\\
0& \frac{1-r}{2}
\end{pmatrix}
~.
\end{equation}
Therefore, the total charging energy can be estimated as
\begin{align}\label{ChEnergy}
\Delta E_{B}&=E(\tau)-E(0)\nonumber\\
&=\frac{\omega_{0}}{2}\Big(1+r_3 (\tau)\Big)-\frac{\alpha}{2}\Big(1+r_3 (0)\Big)\nonumber\\
&=\frac{\omega_{0}}{2}\Big(1+r_3(\tau)\Big)~.
\end{align}
The energy associated to the passive state and active state becomes
\begin{align}
E_{\sigma}(\tau)&=Tr[\sigma_{s}(\tau) H_{B}(\tau)]=\frac{\omega_{0}}{2}\Big(1-r(\tau)\Big)\,\\
E_{\pi}(\tau)&=Tr[\pi_{s}(\tau) H_{B}(\tau)]=\frac{\omega_{0}}{2}\Big(1+r(\tau)\Big)~.
\end{align}
Hence, the ergotropy of the battery can be obtained as
\begin{align}\label{Ergo}
\mathcal{E}&=E(\tau)-E_{\sigma}(\tau)\nonumber\\
&=\frac{\omega_{0}}{2}\Big(r (\tau)+r_3 (\tau)\Big)\,.
\end{align}
Similarly, the antiergotropy of the battery can be obtained as
\begin{align}\label{AntiErgo}
\mathscr{A}&=E(\tau)-E_{\pi}(\tau)
\nonumber\\
&=\frac{\omega_{0}}{2}\Big(r_3 (\tau)-r (\tau)\Big)\,.
\end{align}
Using the values of ergotropy and net charging energy, charging efficiency of the battery takes the form
\begin{equation}
\eta=\frac{r (\tau)+r_3 (\tau)}{1+r_3 (\tau)}\,.
\end{equation}
From the values of ergotropy and antiergotropy, the capacity of the battery becomes
\begin{align}\label{Capacity}
\mathcal{C}&=\frac{\omega_{0}}{2}\Big(r (\tau)+r_3 (\tau)\Big)-\frac{\omega_{0}}{2}\Big(r_3 (\tau)-r (\tau)\Big)\nonumber\\
&=\omega_{0} r(\tau)\,.
\end{align}
\section{Relativistic effects on the Unruh-DeWitt battery parameters}\label{sec:Rel_effect}
\noindent In this section, we investigate the impact of relativistic effects on the performance of the Unruh-DeWitt quantum battery under two distinct coupling regimes: linear and quadratic.
\subsection{Linear Coupling}
\noindent In this case, we use the expressions for the coefficients $\mathcal{A},\,\mathcal{B}$, and $\mathcal{C}$ for the non-relativistic limit. These expressions are given in Appendix \ref{app:A} in eq.~\eqref{ABC_Nonrelativistic}. Using eq.~\eqref{ABC_Nonrelativistic} in eq.(s) (\ref{l_Bloch}, \ref{l_Bloch_Norm}), the components of the Bloch vector and its norm can be recast as

\begin{align}\label{BlochN}
\vec{r}_{lin}(\mathfrak{t})=&\left(J^{-1}(e^{-2\mathfrak{t}J}-1),\,\,e^{-\mathfrak{t}(J+2K)}\sin(16\pi\mathfrak{t}/\lambda^2),\,\,-e^{-\mathfrak{t}(J+2K)}\cos(16\pi\mathfrak{t}/\lambda^2) \right)\,
\end{align}
and
\begin{equation}\label{BlochN_Norm}
r_{lin}\equiv\vert \vec{r}_{lin}(\mathfrak{t})\vert=\sqrt{J^{-2}(e^{-2\mathfrak{t}J}-1)^2 +e^{-2\mathfrak{t}(J+2K)}}\,
\end{equation}
where 
\begin{eqnarray}
    J &\equiv J(\mathfrak{a},\mathcal{u}) =\left[\coth\left(\frac{\pi}{\mathfrak{a}}\right)-4\pi\mathcal{h}(\mathfrak{a})\mathcal{u}^2\right]
\end{eqnarray}
\begin{eqnarray}
    K &\equiv K(\mathfrak{a},\mathcal{u}) =\frac{\mathfrak{a}}{\pi}\left[1-\left(\frac{7}{6}-\frac{\pi^2}{9}\right)\mathcal{u}^2\right]
\end{eqnarray}
with 
\begin{eqnarray}
    \mathcal{h}(\mathfrak{a})=\frac{\mathfrak{a}e^{2\pi/ \mathfrak{a}}}{6(e^{2\pi/ \mathfrak{a}}-1)^2}\left[2+\frac{9}{\mathfrak{a}^2}-\frac{2\pi}{\mathfrak{a}}\left(1+\frac{1}{\mathfrak{a}^2}\right)\coth\left(\frac{\pi}{\mathfrak{a}}\right)\right]~.\nonumber\\
\end{eqnarray}
Therefore, using eq.(s)~(\ref{BlochN}, \ref{BlochN_Norm}), all the quantum battery parameters in the non-relativistic limit for linear field coupling take the form
\begin{equation}
    \mathsf{E}\equiv\frac{\mathcal{E}}{\omega_{0}}=\frac{1}{2}\left[\sqrt{J^{-2}(e^{-2\mathfrak{t}J}-1)^2 +e^{-2\mathfrak{t}(J+2K)}}-e^{-\mathfrak{t}(J+2K)}\cos(16\pi\mathfrak{t}/\lambda^2)\right]
\end{equation}
\begin{equation}
    \mathsf{C}\equiv\frac{\mathcal{C}}{\omega_{0}}=\sqrt{J^{-2}(e^{-2\mathfrak{t}J}-1)^2+e^{-2\mathfrak{t}(J+2K)}}
\end{equation}
\begin{equation}
\eta=\frac{\left[\sqrt{J^{-2}(e^{-2\mathfrak{t}J}-1)^2 +e^{-2\mathfrak{t}(J+2K)}}-e^{-\mathfrak{t}(J+2K)}\cos(16\pi\mathfrak{t}/\lambda^2)\right]}{\Big[1-e^{-\mathfrak{t}(J+2K)}\cos(16\pi\mathfrak{t}/\lambda^2)\Big]}~.
\end{equation}
Similarly, using the expressions for the coefficients $\mathcal{A},\,\mathcal{B}$, and $\mathcal{C}$ for the relativistic limit, given in Appendix \ref{app:A} in eq.~\eqref{ABC_Ultrarelativistic} into eq.(s)~(\ref{l_Bloch}, \ref{l_Bloch_Norm}), the components of the Bloch vector and its norm takes the following form
\begin{align}\label{BlochU}
\vec{r}_{lin}(\mathfrak{t})=&\left(L^{-1}\left(e^{-2\mathfrak{t}L\mathcal{u}^{-2}}-1\right),\,\,e^{-\mathfrak{t}\left(L\mathcal{u}^{-2}+2M\right)}\sin(16\pi\mathfrak{t}/\lambda^2),\,\,-e^{-\mathfrak{t}\left(L\mathcal{u}^{-2}+2M\right)}\cos(16\pi\mathfrak{t}/\lambda^2)\right)\,
\end{align}
and
\begin{equation}\label{BlochU_Norm}
r_{lin}=\sqrt{L^{-2}\left(e^{-2\mathfrak{t}L\mathcal{u}^{-2}}-1\right)^2 +e^{-2\mathfrak{t}\left(L\mathcal{u}^{-2}+2M\right)}}\,
\end{equation}
where 
\begin{equation}
L \equiv L(\mathfrak{a},\mathcal{u}) =\coth\left(\frac{\pi\mathcal{u}}{\mathfrak{a}}\right)\,,\hspace{0.5cm}
M \equiv M(\mathfrak{a},\mathcal{u}) =\frac{\mathfrak{a}}{\pi\mathcal{u}^3}\,.
\end{equation}
Therefore, using eq.(s)~(\ref{BlochU}, \ref{BlochU_Norm}), all the quantum battery parameters for linear field coupling in the relativistic limit turn out to be
\begin{equation}
    \mathsf{E}=\frac{1}{2}\left[\sqrt{L^{-2}\left(e^{-2\mathfrak{t}L\mathcal{u}^{-2}}-1\right)^2 +e^{-2\mathfrak{t}\left(L\mathcal{u}^{-2}+2M\right)}}-e^{-\mathfrak{t}\left(L\mathcal{u}^{-2}+2M\right)}\cos(16\pi\mathfrak{t}/\lambda^2)\right]
\end{equation}
\begin{equation}
\mathsf{C}=\sqrt{L^{-2}\left(e^{-2\mathfrak{t}L\mathcal{u}^{-2}}-1\right)^2 +e^{-2\mathfrak{t}\left(L\mathcal{u}^{-2}+2M\right)}}
\end{equation}
\begin{equation}
\eta=\frac{\left[\sqrt{L^{-2}\left(e^{-2\mathfrak{t}Lu^{-2}}-1\right)^2 +e^{-2\mathfrak{t}\left(Lu^{-2}+2M\right)}}-e^{-\mathfrak{t}\left(Lu^{-2}+2M\right)}\cos(16\pi\mathfrak{t}/\lambda^2)\right]}{\Big[1- e^{-\mathfrak{t}\left(Lu^{-2}+2M\right)}\cos(16\pi\mathfrak{t}/\lambda^2)\Big]}~.
\end{equation}

\noindent In the above equation, the dimensionless parameters are given by 
$$\mathfrak{t}=\frac{\lambda^2\alpha\tau}{16\pi}\,\,,\hspace{0.5cm} \mathfrak{a}=\frac{a}{\alpha}~.$$

\subsection{Quadratic Coupling}
\noindent Now, we move to the case of the quadratic scalar field coupling. Using the expressions for the coefficients $\mathcal{R},\,\mathcal{S}$, and $\mathcal{T}$ for the non-relativistic limit, given in eq.~\eqref{ABC1A_Nonrelativistic} into eq.(s) (\ref{q_Bloch}, \ref{q_Bloch_Norm}), the components of the Bloch vector and its norm can be recast as

\begin{align}\label{Bloch1AN}
\vec{r}_{quad}(\mathfrak{t}_1)=&\bigg(P^{-1}(e^{-2\mathfrak{t}_{1}(1+\mathfrak{a}^2)P}-1),\,\,e^{-\mathfrak{t}_{1}((1+\mathfrak{a}^2)P+2Q)}\sin(192\pi^{3}\mathfrak{t}_{1}/\mu^2\alpha^2),\nonumber\\
&\,\,-e^{-\mathfrak{t}_{1}((1+\mathfrak{a}^2)P+2Q)}\cos(192\pi^{3}\mathfrak{t}_{1}/\mu^2\alpha^2) \bigg)
\end{align}
and
\begin{equation}\label{Bloch1AN_Norm}
r_{quad}=\sqrt{P^{-2}(e^{-2\mathfrak{t}_{1}(1+\mathfrak{a}^2)P}-1)^2 +e^{-2\mathfrak{t}_{1}((1+\mathfrak{a}^2)P+2Q)}}\,
\end{equation}
where 
\begin{align}
P &\equiv P(\mathfrak{a},\mathcal{u}) =\left[\coth\left(\frac{\pi}{\mathfrak{a}}\right)-2\pi\mathcal{j}(\mathfrak{a})\mathcal{u}^2\right]\,,\\
Q &\equiv Q(\mathfrak{a},\mathcal{u}) =\frac{\mathfrak{a}^3}{\pi}\left[1-\left(\frac{5}{2}-\frac{4\pi^2}{15}\right)\mathcal{u}^2\right]
\end{align}
with
\begin{align}
\mathcal{j}(\mathfrak{a})=&\frac{\mathfrak{a}^{3}e^{2\pi/ \mathfrak{a}}}{5(1+\mathfrak{a}^2)(e^{2\pi/ \mathfrak{a}}-1)^2}\left[8+\frac{35}{\mathfrak{a}^2}+\frac{15}{\mathfrak{a}^4}-\frac{2\pi}{\mathfrak{a}}\left(4+\frac{5}{\mathfrak{a}^2}+\frac{1}{\mathfrak{a}^4}\right)\coth\left(\frac{\pi}{\mathfrak{a}}\right)\right]~.
\end{align}
Therefore, using eq.(s)~(\ref{Bloch1AN}, \ref{Bloch1AN_Norm}), all the quantum battery parameters in the non-relativistic limit for the quadratic scalar field coupling take the form
\begin{equation}
    \mathsf{E}=\frac{1}{2}\left[\sqrt{P^{-2}(e^{-2\mathfrak{t}_{1}(1+\mathfrak{a}^2)P}-1)^2 +e^{-2\mathfrak{t}_{1}((1+\mathfrak{a}^2)P+2Q)}}-e^{-\mathfrak{t}_{1}((1+\mathfrak{a}^2)P+2Q)}\cos(192\pi^{3}\mathfrak{t}_{1}/\mu^2\alpha^2)\right]
\end{equation}
\begin{equation}
    \mathsf{C}=\sqrt{P^{-2}(e^{-2\mathfrak{t}_{1}(1+\mathfrak{a}^2)P}-1)^2 +e^{-2\mathfrak{t}_{1}((1+\mathfrak{a}^2)P+2Q)}}
\end{equation}
\begin{equation}
\eta=\frac{\left[\sqrt{P^{-2}(e^{-2\mathfrak{t}_{1}(1+\mathfrak{a}^2)P}-1)^2 +e^{-2\mathfrak{t}_{1}((1+\mathfrak{a}^2)P+2Q)}}--e^{-\mathfrak{t}_{1}((1+\mathfrak{a}^2)P+2Q)}\cos(192\pi^{3}\mathfrak{t}_{1}/\mu^2\alpha^2)\right]}{\Big[1- e^{-\mathfrak{t}_{1}((1+\mathfrak{a}^2)P+2Q)}\cos(192\pi^{3}\mathfrak{t}_{1}/\mu^2\alpha^2)\Big]}~.
\end{equation}
Similarly, using the expressions for the coefficients $\mathcal{R},\,\mathcal{S}$, and $\mathcal{T}$ for the relativistic limit, given in eq.~\eqref{ABC1A_Ultrarelativistic} into eq.(s)~(\ref{q_Bloch}, \ref{q_Bloch_Norm}), the components of the Bloch vector and its norm takes the following form
\begin{align}\label{Bloch1AU}
\vec{r}_{quad}(\mathfrak{t}_{1})=&\bigg(L^{-1}\left(e^{-2\mathfrak{t}_{1}\left(1+\mathfrak{a}^2\mathcal{u}^{-2}\right)L\mathcal{u}^{-4}}-1\right),\,\,e^{-\mathfrak{t}_{1}\big(\left(1+\mathfrak{a}^2\mathcal{u}^{-2}\right)L\mathcal{u}^{-4}+2V\big)}\sin(192\pi^{3}\mathfrak{t}_{1}/\mu^2\alpha^2),\nonumber\\
&\,\,-e^{-\mathfrak{t}_{1}\big(\left(1+\mathfrak{a}^2\mathcal{u}^{-2}\right)L\mathcal{u}^{-4}+2V\big)}\cos(192\pi^{3}\mathfrak{t}_{1}/\mu^2\alpha^2)\bigg)
\end{align}
and
\begin{align}\label{Bloch1AU_Norm}
r_{quad}=&\left[L^{-2}\left(e^{-2\mathfrak{t}_{1}\left(1+\mathfrak{a}^2\mathcal{u}^{-2}\right)L\mathcal{u}^{-4}}-1\right)^2+e^{-2\mathfrak{t}_{1}\big(\left(1+\mathfrak{a}^2\mathcal{u}^{-2}\right)L\mathcal{u}^{-4}+2V\big)}\right]^{1/2}
\end{align}
where 
\begin{equation}
L = L(\mathfrak{a},\mathcal{u}) =\coth\left(\frac{\pi\mathcal{u}}{\mathfrak{a}}\right)\,,\hspace{0.5cm}
V \equiv V(\mathfrak{a},\mathcal{u}) =\frac{\mathfrak{a}^3}{\pi\mathcal{u}^7}\,.
\end{equation}
Therefore, using eq.(s)~(\ref{Bloch1AU}, \ref{Bloch1AU_Norm}), all the quantum battery parameters in the ultrarelativistic limit for the quadratic scalar field coupling turn out to be
\begin{align}
\mathsf{E}&=\frac{1}{2}\Bigg[\left[L^{-2}\left(e^{-2\mathfrak{t}_{1}\left(1+\frac{\mathfrak{a}^2}{\mathcal{u}^{2}}\right)\frac{L}{\mathcal{u}^{4}}}-1\right)^2+e^{-2\mathfrak{t}_{1}\Big(\left(1+\frac{\mathfrak{a}^2}{\mathcal{u}^{2}}\right)\frac{L}{\mathcal{u}^{4}}+2V\Big)}\right]^{1/2}-e^{-\mathfrak{t}_{1}\Big(\left(1+\frac{\mathfrak{a}^2}{\mathcal{u}^{2}}\right)\frac{L}{\mathcal{u}^{4}}+2V\Big)}\cos\left(\frac{192\pi^{3}\mathfrak{t}_{1}}{\mu^2\alpha^2}\right)\Bigg]
\end{align}
\begin{align}
\mathsf{C}&=\left[L^{-2}\left(e^{-2\mathfrak{t}_{1}\left(1+\frac{\mathfrak{a}^2}{\mathcal{u}^{2}}\right)\frac{L}{\mathcal{u}^{4}}}-1\right)^2+e^{-2\mathfrak{t}_{1}\Big(\left(1+\frac{\mathfrak{a}^2}{\mathcal{u}^{2}}\right)\frac{L}{\mathcal{u}^{4}}+2V\Big)}\right]^{1/2}
\end{align}
\begin{align}
\eta=&
\Bigg[\left[L^{-2}\left(e^{-2\mathfrak{t}_{1}\left(1+\frac{\mathfrak{a}^2}{\mathcal{u}^{2}}\right)\frac{L}{\mathcal{u}^{4}}}-1\right)^2+e^{-2\mathfrak{t}_{1}\Big(\left(1+\frac{\mathfrak{a}^2}{\mathcal{u}^{2}}\right)\frac{L}{\mathcal{u}^{4}}+2V\Big)}\right]^{1/2}-e^{-\mathfrak{t}_{1}\Big(\left(1+\frac{\mathfrak{a}^2}{\mathcal{u}^{2}}\right)\frac{L}{\mathcal{u}^{4}}+2V\Big)}\cos\left(\frac{192\pi^{3}\mathfrak{t}_{1}}{\mu^2\alpha^2}\right)\Bigg]\nonumber\\
&\times\Bigg[1-e^{-\mathfrak{t}_{1}\Big(\left(1+\frac{\mathfrak{a}^2}{\mathcal{u}^{2}}\right)\frac{L}{\mathcal{u}^{4}}+2V\Big)}\cos\left(\frac{192\pi^{3}\mathfrak{t}_{1}}{\mu^2\alpha^2}\right)\Bigg]^{-1} ~.
\end{align}
In the above equation, the dimensionless parameters are given by $$\mathfrak{t}_{1}=\frac{\mu^2\alpha^{3}\tau}{192\pi^3}\,\,,\hspace{0.8cm}\mathfrak{a}=\frac{a}{\alpha}~.$$  


\section{Results and discussion}\label{sec:Results}

\noindent We are now ready to discuss our results obtained for the key battery parameters under quadratic coupling to the scalar field environment and to compare their behavior across different coupling scenarios.

\subsection{Behaviour of the quantum battery parameters with respect to acceleration}
\begin{figure}[!ht]
 		\centering
 		\begin{subfigure}{0.45\textwidth}
 			\centering
 			\includegraphics[width=\textwidth]{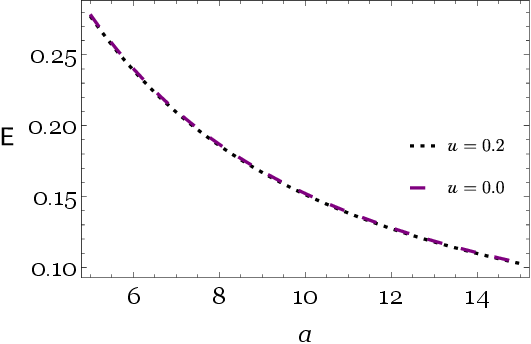}
 			\caption{Behaviour of ergotropy in the non-relativistic limit.}
 			\label{fig:ErQNR}
 		\end{subfigure}
 		\hfill
 		\begin{subfigure}{0.45\textwidth}
 			\centering
 			\includegraphics[width=\textwidth]{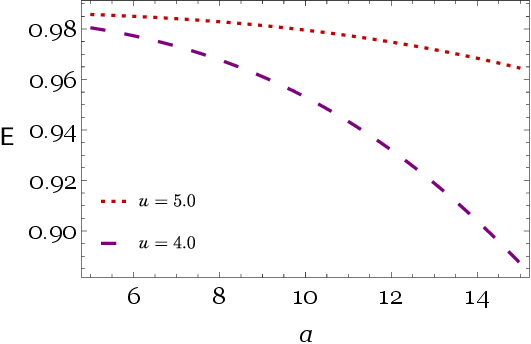}
 			\caption{Behaviour of ergotropy in the relativistic limit.}
 			\label{fig:ErQUR}
 		\end{subfigure}
 		\newline
 		\begin{subfigure}{0.45\textwidth}
 			\centering
 			\includegraphics[width=\textwidth]{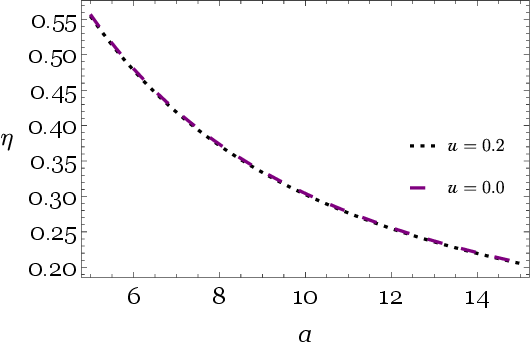}
 			\caption{Behaviour of charging efficiency in the non-relativistic limit.}
 			\label{fig:EfQNR}
 		\end{subfigure}\hfill
 		\begin{subfigure}{0.45\textwidth}
 			\centering
 			\includegraphics[width=\textwidth]{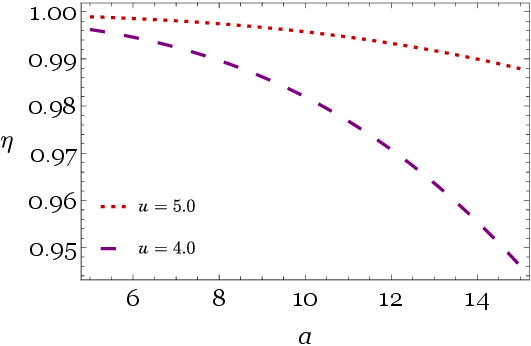}
 			\caption{Behaviour of charging efficiency in the relativistic limit.}
 			\label{fig:EfQUR}
 		\end{subfigure}
 		\newline
 		\begin{subfigure}{0.45\textwidth}
 			\centering
 			\includegraphics[width=\textwidth]{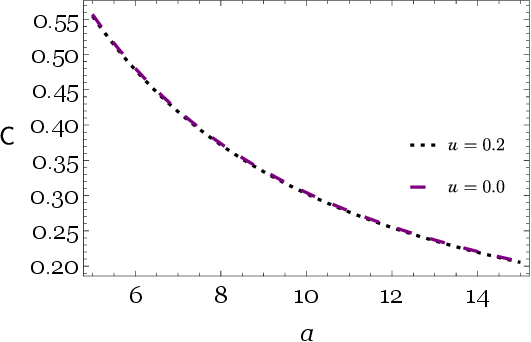}
 			\caption{Behaviour of capacity in the non-relativistic limit.}
 			\label{fig:CaQNR}
 		\end{subfigure}\hfill
 		\begin{subfigure}{0.45\textwidth}
 			\centering
 			\includegraphics[width=\textwidth]{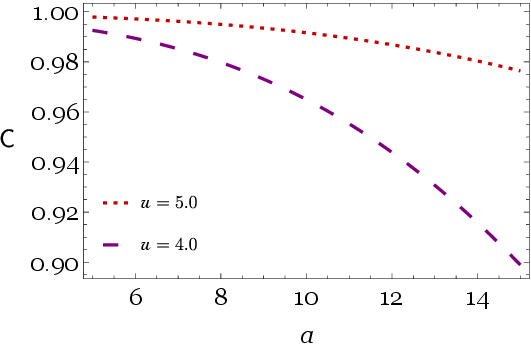}
 			\caption{Behaviour of capacity in the relativistic limit.}
 			\label{fig:CaQUR}
 		\end{subfigure}
 		\caption{Behaviour of different battery parameters with respect to $a$ in quadratic coupling scenario.}
 		\label{fig:fig1}
 	\end{figure}

\noindent In Fig. \ref{fig:fig1}, the effect of acceleration on various quantum battery parameters is depicted for both the non-relativistic and relativistic limits. 
 \begin{figure}[!ht]
 	\centering
 	\begin{subfigure}{0.45\textwidth}
 		\centering
 		\includegraphics[width=\textwidth]{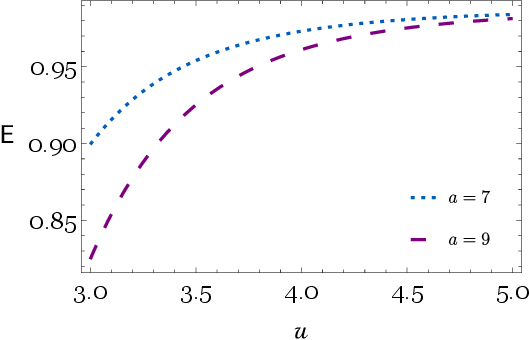}
 		\caption{Behaviour of ergotropy.}
 		\label{fig:ErVQUR}
 	\end{subfigure}\hfill
 	\begin{subfigure}{0.45\textwidth}
 		\centering
 		\includegraphics[width=\textwidth]{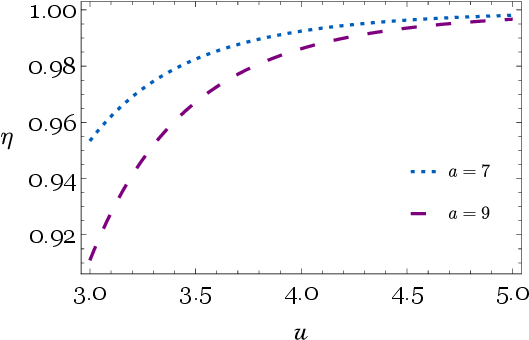}
 		\caption{Behaviour of charging efficiency.}
 		\label{fig:EfVQUR}
 	\end{subfigure}
 	\newline
 	\begin{subfigure}{0.45\textwidth}
 		\centering
 		\includegraphics[width=\textwidth]{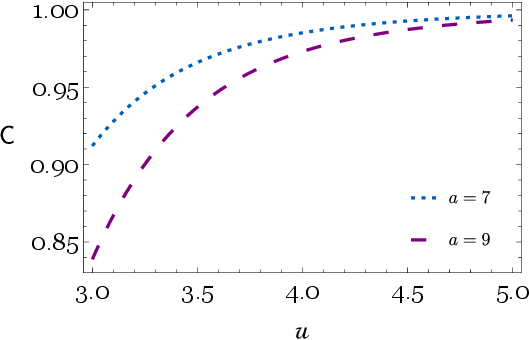}
 		\caption{Behaviour of capacity.}
 		\label{fig:CaVQUR}
 	\end{subfigure}\hfill
 	\caption{Behaviour of different battery parameters with respect to $\mathcal{u}$ in the relativistic limit.}
 	\label{fig:fig2}
 \end{figure}
 \begin{figure}[!t]
 	\centering
 	\begin{subfigure}{0.45\textwidth}
 		\centering
 		\includegraphics[width=\textwidth]{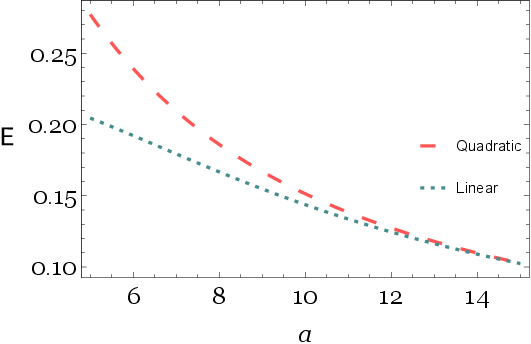}
 		\caption{Behaviour of ergotropy in the non-relativistic limit, $\mathcal{u}=0.2$.}
 		\label{fig:ErCNR}
 	\end{subfigure}\hfill
 	\begin{subfigure}{0.45\textwidth}
 		\centering
 		\includegraphics[width=\textwidth]{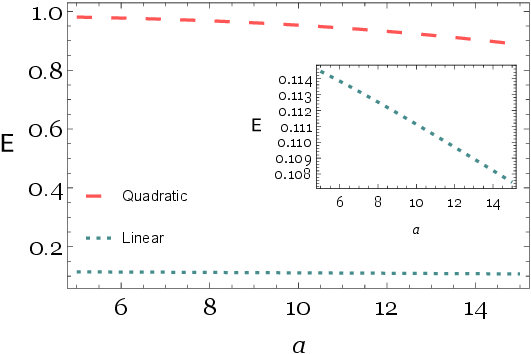}
 		\caption{Behaviour of ergotropy in the relativistic limit, $\mathcal{u}=4$.}
 		\label{fig:ErCUR}
 	\end{subfigure}
 	\newline
 	\begin{subfigure}{0.45\textwidth}
 		\centering
 		\includegraphics[width=\textwidth]{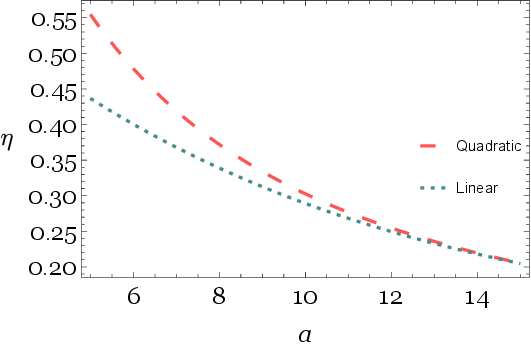}
 		\caption{Behaviour of charging efficiency in the non-relativistic limit, $\mathcal{u}=0.2$.}
 		\label{fig:EfCNR}
 	\end{subfigure}\hfill
 	\begin{subfigure}{0.45\textwidth}
 		\centering
 		\includegraphics[width=\textwidth]{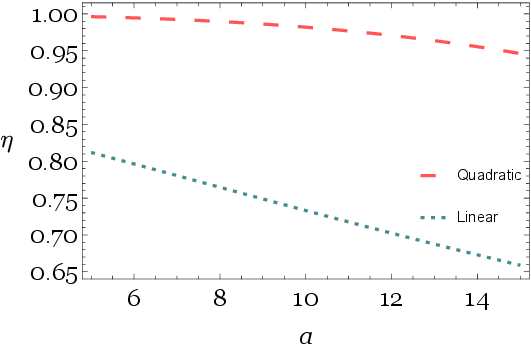}
 		\caption{Behaviour of charging efficiency in the relativistic limit, $\mathcal{u}=4$.}
 		\label{fig:EfCUR}
 	\end{subfigure}
 	\newline
 	\begin{subfigure}{0.45\textwidth}
 		\centering
 		\includegraphics[width=\textwidth]{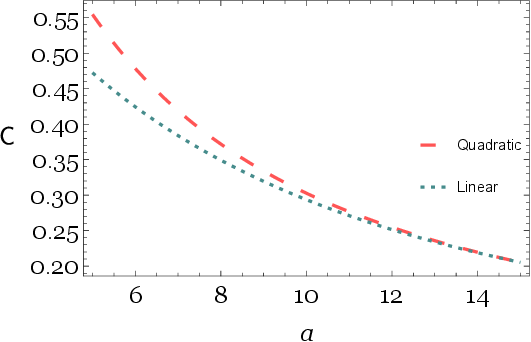}
 		\caption{Behaviour of capacity in the non-relativistic limit, $\mathcal{u}=0.2$.}
 		\label{fig:CaCNR}
 	\end{subfigure}\hfill
 	\begin{subfigure}{0.45\textwidth}
 		\centering
 		\includegraphics[width=\textwidth]{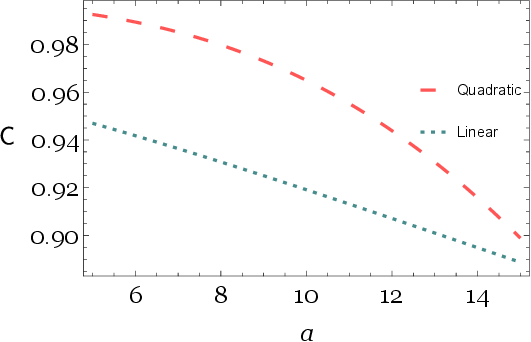}
 		\caption{Behaviour of capacity in the relativistic limit, $\mathcal{u}=4$.}
 		\label{fig:CaCUR}
 	\end{subfigure}
 	\caption{Behaviour of different battery parameters with respect to $a$ in different coupling scenarios.}
 	\label{fig:fig3}
 \end{figure}
  \begin{figure}[!ht]
 	\centering
 	\begin{subfigure}{0.45\textwidth}
 		\centering
 		\includegraphics[width=\textwidth]{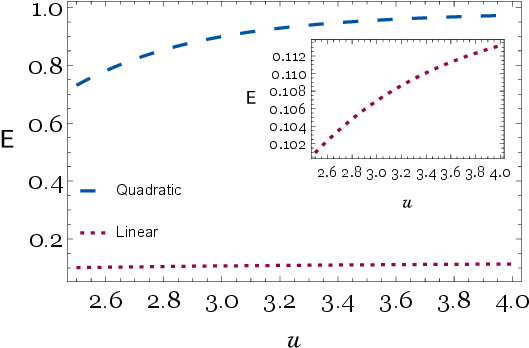}
 		\caption{Behaviour of ergotropy.}
 		\label{fig:ErCVUR}
 	\end{subfigure}\hfill
 	\begin{subfigure}{0.45\textwidth}
 		\centering
 		\includegraphics[width=\textwidth]{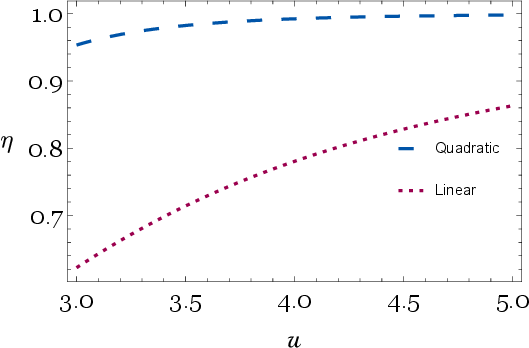}
 		\caption{Behaviour of charging efficiency.}
 		\label{fig:EfCVUR}
 	\end{subfigure}
 	\newline
 	\begin{subfigure}{0.45\textwidth}
 		\centering
 		\includegraphics[width=\textwidth]{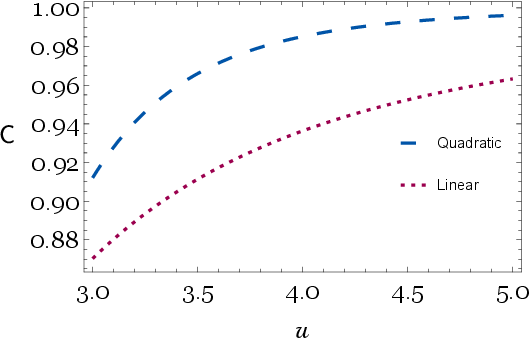}
 		\caption{Behaviour of capacity.}
 		\label{fig:CaCVUR}
 	\end{subfigure}\hfill
 	\caption{Behaviour of different battery parameters with respect to $\mathcal{u}$ in the relativistic limit in different coupling scenario.}
 	\label{fig:fig4}
 \end{figure}
For both the limiting case, we choose two different $\mathcal{u}$ values. For the entire analysis values of $\mathcal{u}$ can be fixed using the relation $\mathcal{u}=\frac{\mathit{v}}{\sqrt{1-\mathit{v}^2}}$, where $\mathit{v}$ is the magnitude of the battery velocity. In the following Figures, we consider all the battery parameters and velocity $\mathcal{u}$ as dimensionless parameters. Other than this, we consider the dimensionless parameters $\mathfrak{a}$, $\mathfrak{t}_{1}$ as $a$, and $\tau$ respectively. In the following figures we set $\tau=0.5$.

The variation of scaled ergotropy across non-relativistic and relativistic regimes is showcased in Figures \ref{fig:ErQNR} and \ref{fig:ErQUR}, respectively. From Fig. \ref{fig:ErQNR}, we observe that for a stationary battery ($\mathit{v} = 0$, $\mathcal{u} = 0$), the ergotropy steadily decreases with increasing acceleration. This decline is attributed to decoherence induced by the interaction between the battery and the massless scalar field. When introducing a non-zero velocity, we find that, for a fixed acceleration, variations in velocity do not significantly affect the ergotropy. Thus, in the non-relativistic regime, velocity fails to mitigate the degradation in battery performance caused by acceleration.

In contrast, Fig. \ref{fig:ErQUR} illustrates the relativistic regime, where the battery velocities are $\mathit{v} = 0.97$ ($\mathcal{u} = 4$) and $\mathit{v} = 0.99$ ($\mathcal{u} = 5$). Although ergotropy still decreases with increasing acceleration, the rate of decrease is substantially reduced compared to the non-relativistic case. Moreover, at a fixed acceleration, increasing the velocity from $\mathit{v} = 0.97$ to $0.99$ ($\mathcal{u} = 4\rightarrow5$) leads to a notable enhancement in ergotropy. This demonstrates that in the relativistic limit, the inclusion of velocity effects can significantly improve the performance of the quantum battery.

Figures \ref{fig:EfQNR} and \ref{fig:EfQUR} depict the variation of the charging efficiency of the quantum battery in the non-relativistic and relativistic regimes, respectively. Similarly, Figures \ref{fig:CaQNR} and \ref{fig:CaQUR} show the corresponding behaviour of the battery's capacity in these two limits. In both regimes, the charging efficiency and capacity exhibit a similar trend of decreasing with increasing acceleration. In the non-relativistic case, increasing the battery's velocity has negligible impact on either parameter at a fixed acceleration. In contrast, in the relativistic regime, the adverse effects of decoherence are mitigated by the influence of the four-velocity. As a result, for a fixed acceleration, increasing the battery's velocity leads to a marked improvement in both charging efficiency and capacity.
\subsection{Behaviour of the quantum battery parameters with respect to velocity}
\noindent In Fig. \ref{fig:fig2}, we illustrate how the velocity of the battery influences various quantum battery parameters in the relativistic regime. Specifically, we analyse the behaviour by choosing two distinct values of acceleration, \( a = 7 \) and \( a = 9 \), to examine how increased acceleration interacts with relativistic motion.

Figure \ref{fig:ErVQUR} presents the variation of the scaled ergotropy as a function of the dimensionless velocity parameter \( \mathcal{u} \) in the relativistic limit. As previously observed in Fig. \ref{fig:fig1}, the quantum battery parameters remain largely unaffected by changes in velocity in the non-relativistic regime. Therefore, in this discussion, we focus solely on the relativistic limit, where such variations become prominent.

From Fig. \ref{fig:ErVQUR}, we find that for a fixed value of velocity, increasing the battery’s acceleration from \( a = 7 \) to \( a = 9 \) leads to a noticeable decline in the ergotropy. This trend aligns with the implications of the Unruh effect as acceleration increases, the detector experiences a thermal environment with a higher effective temperature. This thermal noise introduces additional fluctuations, which in turn cause decoherence by randomising the system’s quantum state and diminishing its coherence.
Interestingly, we also observe that the ergotropy grows substantially with increasing velocity. This suggests that relativistic motion plays a constructive role in enhancing the battery's performance. However, even at higher velocities, an increase in acceleration still leads to a reduction in ergotropy, reaffirming the disruptive role of thermal noise associated with the Unruh effect. Figures \ref{fig:EfVQUR} and \ref{fig:CaVQUR} display the corresponding variations in the charging efficiency and the energy capacity of the battery, respectively, within the relativistic regime. Both of these parameters exhibit trends similar to that of ergotropy: they increase significantly with rising velocity and decrease with higher acceleration at fixed velocity. Overall, across all three battery parameters becomes evident that introducing motion in a direction orthogonal to the acceleration suppresses the decoherence effects induced by the thermal environment. As a result, the presence of relativistic velocity substantially enhances the operational capabilities of the quantum battery, offering a promising avenue for optimising performance through controlled kinematics.
\subsection{Comparison between the linear and quadratic couplings}
\noindent To compare the effects of linear and quadratic environmental coupling on the performance of the quantum battery, we examine the three key battery parameters in both the non-relativistic and relativistic regimes. In Fig. \ref{fig:fig3}, we present how these parameters vary with acceleration \( a \) under the two coupling scenarios across both regimes.

Figure \ref{fig:ErCNR} focuses on the behaviour of the scaled ergotropy in the non-relativistic regime. For a fixed value of the battery's velocity, we observe that the ergotropy decreases monotonically with increasing acceleration, irrespective of whether the coupling is linear or quadratic. Notably, the curves corresponding to linear and quadratic couplings tend to converge as the acceleration becomes large. Furthermore, for any lower value of acceleration, the scaled ergotropy is initially higher in the case of quadratic coupling compared to its linear counterpart, indicating an enhanced ability to store extractable work in the presence of nonlinear interactions.

From Fig. \ref{fig:ErCUR}, which presents the results in the relativistic regime, we observe that for a fixed value of the quantum battery velocity, the ergotropy decreases only mildly with increasing acceleration in the case of linear coupling. In contrast, for quadratic coupling, the ergotropy remains nearly constant, showing negligible dependence on acceleration. This behaviour suggests that when the battery velocity is sufficiently high (e.g., \( \mathit{v} \sim 0.97 \)), quadratic coupling effectively suppresses decoherence, and the quantum battery behaves nearly as a closed system. Such coherence-preserving dynamics have also been observed in cavity-based environments \cite{Yao_2015}. The preservation of quantum coherence is vital for the optimal functioning of quantum devices such as batteries.

Figures \ref{fig:EfCNR} and \ref{fig:EfCUR} display the charging efficiency in the non-relativistic and relativistic regimes, respectively. From Fig. \ref{fig:EfCNR}, we see that for a fixed battery velocity in the non-relativistic regime, both linear and quadratic couplings yield qualitatively similar trends to that of ergotropy: the charging efficiency decreases with increasing acceleration. At higher accelerations, the distinction between the two coupling types becomes less pronounced. In the relativistic regime, as shown in Fig. \ref{fig:EfCUR}, the efficiency under linear coupling drops more significantly with acceleration. However, for quadratic coupling, the decline is much more gradual. Moreover, for fixed values of \( a \) and \( \mathcal{u} \), the charging efficiency is consistently higher under quadratic coupling. This trend highlights that the performance enhancement provided by quadratic coupling becomes even more pronounced in the relativistic regime.

Figures \ref{fig:CaCNR} and \ref{fig:CaCUR} illustrate the behaviour of the battery capacity in the non-relativistic and relativistic regimes, respectively. In the non-relativistic case (Fig. \ref{fig:CaCNR}), the battery capacity follows a similar pattern to the other parameters: it is higher under quadratic coupling, particularly at low acceleration, where decoherence is less significant. As acceleration increases, the influence of the coupling type on the capacity diminishes. However, in the relativistic regime (Fig. \ref{fig:CaCUR}), the advantage of quadratic coupling becomes more prominent. Across all acceleration values, the battery capacity under quadratic coupling is substantially higher than that under linear coupling. This demonstrates that coupling the quantum battery quadratically to the scalar field environment can significantly enhance its overall performance in relativistic settings.

Figure \ref{fig:fig4} exhibits a trend consistent with that observed in Fig. \ref{fig:fig3}, highlighting the behaviour of the three key quantum battery parameters. Specifically, Figs. \ref{fig:ErCVUR}, \ref{fig:EfCVUR}, and \ref{fig:CaCVUR} show the variations in the ergotropy, charging efficiency, and battery capacity, respectively, within the relativistic regime. Across all three parameters, a clear and consistent pattern emerges: each quantity increases with the velocity parameter \( \mathcal{u} \), and the values attained under quadratic coupling are significantly higher than those obtained under linear coupling. These results strongly indicate that quadratic coupling to the scalar field environment leads to a substantial enhancement in the overall performance of the quantum battery in relativistic regimes.

\section{Conclusions}\label{sec:Conclusion}
\noindent In this study we investigate the relativistic effects on the performance of a quantum battery given in terms of  the battery parameters, ergotropy, charging efficiency, and capacity. The quantum battery comprises a Unruh-De Witt detector which is charged through a coherent classical pulse \cite{Mojaveri2023, Hao2023, Abdolrahimi2014}. At the time of charging, the battery interacts with the environment of a  massless quantum scalar field.  The trajectory of the battery is composed of uniform acceleration along one direction, combined with constant four-velocity components in the orthogonal plane to the acceleration. To explore the effect of the environment of the quantum battery, here we consider a nonlinear quadratic coupling between the quantum battery and the massless quantum scalar field \cite{Alicki2013, Hummer2016, Sachs2017, Gray2018},
 and compare the results obtained with the linear coupling scenario. To carry out the computation of the various quantities, we first begin by deriving the Lindblad equation for the quadratic coupling case in details. The form of the Lindblad equation for the quadratic coupling case has a close resemblance to the linear coupling case, with different coefficients appearing in the equation.
 
 A generic feature of quantum dynamics is the degradation of quantum resources such as coherence and entanglement in the relativistic arena \cite{Bruschi2014, Matsas, Y_M_Huang2019, Z_Zhao2020, Du2021, Liu2021, Benatti_2004, B_L_Hu2012, Moustos2017, Chatterjee2017, Mukherjee_PRA_2024, Sokolov2020}.
 The Unruh effect implies an effective heat bath for accelerated systems,
 resulting in decoherence of quantum states. This trend is clearly
 observed through our computation of the battery parameters, all of
 which are seen to fall with acceleration. However, the presence of
 velocity components orthogonal to the direction of the acceleration
 enables a reduction in the rate of loss of coherence, as again evident
 through our evaluation of the battery ergodicity, capacity and
 charging efficiency. Such a feature is absent in the non-relativistic
 limit, where it is shown that battery velocity is unable to ward off
 the decohering effects of acceleration.
 
 The role of the quadratic scalar field coupling becomes remarkably
 evident both in the non-relativistic and in the relativistic limit, leading to significant improvement
 in battery performance over that obtained through a linear environmental
 coupling. The magnitude of the battery ergodicity
 for quadratic coupling exceeds that in the case of linear coupling. The quantum battery capacity, taking into account both
 the ergodicity and antiergodicity, encodes the overall figure
 of merit for storing energy \cite{Mir2023}. Our results show that battery capacity is
 much better preserved through quadratic coupling under acceleration.
 
 Similarly, we find that the difference in the charging efficiency
 between the quadratic and linear coupling scenarios grows with
 acceleration, again providing a striking example of enhancement of
 the quantum battery performance through quadratic environmental coupling
 in the relativistic scenario. The analysis contained in the present work inspires related investigations into the behaviour of other quantities
 of interest in quantum thermodynamics \cite{Myers_2022} in the presence of nonlinear
 environmental couplings. 
 
 In real physical scenario, this kind of quadratic environmental coupling can be introduced through parametric pumping and other simulation techniques intrinsic to superconducting circuits \cite{ann2022tp}, hybrid
spin-nanomechanical oscillators \cite{wang2016method, munoz2018hybrid}, trapped ions \cite{felicetti2015spectral, cheng2018nonlinear, puebla2019quantum, cong2020selective}, and ultracold atoms \cite{schneeweiss2018cold, dareau2018observation}. Experiments with superconducting artificial atoms have already shown non-dipolar couplings \cite{goetz2018parity}. Moreover, in the upcoming scenario for the development of quantum technologies, the formulation of techniques to preserve quantum coherence through
 environmental engineering is likely to play a key role \cite{Koch_2016, Uchiyama_2018}. 
 Our results should motivate further studies on different types of environmental couplings for quantum heat engines in  relativistic regimes.

\section*{Acknowledgement}
\noindent AM would like to thank Prof. Hongwei Yu for valuable email correspondence, Dr. Ashis Saha, and Dr. Anwesha Chakraborty for insightful discussions. AM acknowledges the \textit{Anusandhan National Research Foundation (ANRF)} for providing an International Travel Grant (Grant No: ITS/2024/004608), which facilitated participation in \textit{Quantum Thermodynamics Down Under 2024} at the \textit{University of Queensland, Brisbane, Australia}, where some of the results of this work were presented. AM also acknowledges support from the \textit{Council of Scientific and Industrial Research (CSIR)} through a Foreign Travel Grant (Grant No: TG/13141/24-HRD), which enabled participation in the \textit{QISS Conference 2025} at the \textit{Institute for Quantum Optics and Quantum Information, Vienna, Austria}, where related findings were presented.
    \begin{appendices}
    \section{Effect of the environment on the  Unruh-DeWitt battery: Linear Coupling Scenario}\label{app:A}
        \noindent In this appendix, we are going to give the form of the \textit{GKSL master equation} by considering the simplest model where the quantum scalar field couples with the detector linearly \cite{DeWitt1979, Unruh1984, birrell1984quantum}. Apart from the master equation, some other results of the linear coupling case is also given. These results will be useful in Section \ref{sec:Rel_effect}. This kind of scenario in the context of quantum battery has been studied in \cite{Hao2023}, where the battery moves in a trajectory composed of uniform acceleration along one direction, combined with constant four-velocity component in the orthogonal direction to the acceleration. In this study we investigate this case by considering a more general trajectory given in eq.~\eqref{trajectory}.

        In the linear coupling scenario, the complete Hamiltonian of the detector-field system in the Interaction picture takes the form
\begin{equation}
\widetilde{H}(\tau)=\widetilde{H}_{B}(\tau)+\widetilde{H}_{F}(\tau)+\widetilde{H}_{I}(\tau)
\end{equation}
where 
\begin{align}
   \widetilde{H}_{B}(\tau)=&H_{B}=\frac{\alpha}{2}(\widetilde{\sigma}_{+}(\tau) +\widetilde{\sigma}_{-}(\tau))\\
    \widetilde{H}_{F}(\tau)=&H_{F}=\int d^{3}\mathbf{k}\,|{\mathbf{k}}|\,a_{\mathbf{k}}^{\dagger}(\tau)a_{\mathbf{k}}(\tau)\\
    \widetilde{H}_{I}(\tau)=&\frac{\lambda}{(2\pi)^{3/2}}\int \frac{d^3\mathbf{k}}{\sqrt{ 2|\mathbf{k}|}}\left(\widetilde{\sigma}_{+}(\tau)\,\hat{a}_{\mathbf{k}} e^{-ik\cdot x(\tau)}+\widetilde{\sigma}_{-}(\tau)\,\hat{a}^{\dagger}_{\mathbf{k}}e^{ik\cdot x(\tau)}\right)\label{lin_Ham}.
\end{align}
Here $\lambda$ is the coupling strength between the battery and the quantum scalar field, which is in general very small.

In this scenario, the reduced density matrix of the UDW battery $\rho_{s}(\tau)$ can be evaluated by solving the \textit{Gorini-Kossakowski-Sudarshan-Lindblad} (GKSL) quantum master equation \cite{GKS_1976, Lindblad1976}
  \begin{equation}\label{l_GKSL}
\frac{\partial \rho_s(\tau)}{\partial \tau}=-i\Big[\mathcal{H}_{eff}, \rho_s(\tau)\Big] + \mathcal{L}\big[\rho_s(\tau)\big]
\end{equation}
where
\begin{eqnarray}\label{H_eff,Lind}
\mathcal{H}_{eff}&=&\frac{1}{2}\left[\alpha+\underbrace{\frac{i\lambda^2}{8}\Big\{\mathcal{K}(-\alpha)-\mathcal{K}(\alpha)\Big\}}_{\omega_{LS}}\right](\sigma_{+}+\sigma_{-})\nonumber\\
&=&\frac{1}{2}\Big[\alpha+\omega_{LS}\Big](\sigma_{+}+\sigma_{-})\nonumber\\
\mathcal{L}\big[\rho_s(\tau)\big]&=&\frac{1}{2}\displaystyle\sum_{i,j=1}^{3} \Delta_{ij}\Big[2\sigma_{j}\rho_s(\tau)\sigma_{i}-\sigma_i\sigma_j\rho_s(\tau)-\rho_s(\tau)\sigma_i\sigma_j\Big]\nonumber\\
\end{eqnarray}
The coefficients of the Kossakowski matrix $\Delta_{ij}$ is given by
\begin{equation}
\Delta_{ij}=\mathcal{A}\delta_{ij}-i\mathcal{B}\varepsilon_{ijk}\delta_{k1}-\mathcal{C}\delta_{i1}\delta_{j1}~
\end{equation}
where 
\begin{eqnarray}\label{A,B,C}
    \mathcal{A}&=&\frac{\lambda^2}{16}\Big[\mathcal{G}(\alpha)+\mathcal{G}(-\alpha)\Big]\,\nonumber\\
    \mathcal{B}&=&\frac{\lambda^2}{16}\Big[\mathcal{G}(\alpha)-\mathcal{G}(-\alpha)\Big]\,\nonumber\\
    \mathcal{C}&=&\frac{\lambda^2}{4}\mathcal{G}(0)-\mathcal{A}~.
\end{eqnarray}
which matches with the result in Ref. \cite{Hao2023} but with different Kossakowski matrix coefficients, $\Delta_{ij}$ as the definitions of $\mathcal{A},\,\mathcal{B}$, and $\mathcal{C}$ are different. At this point, it is important to note that adopting the interaction Hamiltonian in the form presented in Ref. \cite{Hao2023} and without taking the RWA does not lead to the GKSL master equation as stated in that reference. In the standard derivation of the GKSL master equation for a linearly coupled interaction Hamiltonian, the Lamb shift correction $\omega_{LS}$
is typically much smaller than the bare transition frequency $\alpha$, i.e. $\omega_{LS}<<\alpha$, and is therefore often neglected in the literature.      

\noindent In this study we investigate the linear coupling case by considering a more general trajectory given in eq.~\eqref{trajectory}. 

\noindent The Wightman function in the linear coupling takes the form
\begin{align}
&G^{+}(\tau,\tau-\tau')=\langle 0\vert\phi(x(\tau))\phi(x(\tau-\tau'))\vert 0\rangle\nonumber\\
&=-\frac{1}{4\pi^2}\frac{1}{[(t(\tau)-t(\tau-\tau')-i\epsilon)^2-\vert \textbf{x}(\tau)-\textbf{x}(\tau-\tau')\vert^2]}~.
\end{align}
Using the trajectory given in eq.~\eqref{trajectory}, the Wightman function along this trajectory turns out to be
\begin{equation}\label{Wightman}
G^{+}(\tau,\tau-\tau')=-\frac{\xi^4}{16\pi^2 a^2}\left[\sinh^2\left(\frac{\xi\,\tau'}{2}-i\epsilon\right)-\frac{\mathcal{u}^2\xi^4(\tau')^2}{4 a^2}\right]^{-1}~.
\end{equation}
Using the relation $\xi=\frac{a}{\sqrt{1+\mathcal{u}^2}}$ in the above equation, it can be rewritten as
\begin{align}\label{Wightman1}
G^{+}(\tau,\tau-\tau')
&=-\frac{a^2}{16\pi^2 (1+\mathcal{u}^2)^2}\left[\sinh^2\left(\frac{a\,\tau'}{2\sqrt{1+\mathcal{u}^2}}-i\epsilon\right)-\frac{\mathcal{u}^2a^2(\tau')^2}{4(1+\mathcal{u}^2)^2}\right]^{-1}~.
\end{align}
The above result is  more general in the sense that in spite of having a uniform accelaration along the $x$ direction, here we consider two constant velocities in $y$ and $z$ directions. After turning off the $y$ and $z$ directional velocities, it is easily found that $\upsilon$ becomes $a$ and $G^{+}(\tau')$ gives the usual Wightman function for the massless quantum scalar field \cite{Mukherjee2023}.

\noindent We evaluate the Fourier transform of the field correlation function  $\mathcal{G}(\alpha)$ when the battery is moving along the trajectory given by eq.~\eqref{trajectory}. In order to obtain $\mathcal{G}(\alpha)$ analytically,  we consider the following limits enabling us to obtain key physical
insights on our results: (i) $\mathcal{u}^2<<1$ (non-relativistic limit) and (ii) $\mathcal{u}^2>>1$ (relativistic limit). 

\noindent In the limit $\mathcal{u}^2<<1$, the Wightman function upto $\mathcal{O}(\mathcal{u}^2)$ takes the form
\begin{align}\label{Wightman_NR}
G^{+}(\tau,\tau-\tau')=&-\frac{a^2}{16\pi^2}\left[\frac{(1-2\mathcal{u}^2)}{\sinh^2\left(\frac{a(\tau')}{2}-i\epsilon\right)}+\frac{(a(\tau'))^2+a(\tau')\sinh\big(a(\tau')-i\epsilon\big)}{4\sinh^4\left(\frac{a(\tau')}{2}-i\epsilon\right)}\mathcal{u}^2\right]~.
\end{align}
Now calculating $\mathcal{G}(\pm \alpha)$ by using the above Wightman function eq.~\eqref{Wightman_NR}, we get
\begin{equation}\label{g1a1}
    \mathcal{G}(\alpha)=\frac{\alpha}{2\pi}\frac{e^{2\pi\alpha/a}}{(e^{2\pi\alpha/a}-1)}-\mathcal{h}(\alpha, a)\mathcal{u}^2
\end{equation}
\begin{equation}\label{g1a2}
    \mathcal{G}(-\alpha)=\frac{\alpha}{2\pi}\frac{1}{(e^{2\pi\alpha/a}-1)}-\mathcal{h}(\alpha, a)\mathcal{u}^2
\end{equation}
where 
\begin{align}
\mathcal{h}(\alpha, a)
&=\frac{a e^{2\pi\alpha/a}}{6(e^{2\pi\alpha/a}-1)^2}\left[2+\frac{9\alpha^2}{a^2}-\frac{2\pi\alpha}{a}\left(1+\frac{\alpha^2}{a^2}\right)\coth\left(\frac{\pi\alpha}{a}\right)\right].
\end{align}
Taking $\alpha\rightarrow0$ limit, we also get
\begin{equation}\label{g101}
    \mathcal{G}(0)=\frac{a}{4\pi^2}\left[1-\left(\frac{7}{6}-\frac{\pi^2}{9}\right)\mathcal{u}^2\right]~.
\end{equation}
Substituting eq.(s) (\ref{g1a1}, \ref{g1a2}, \ref{g101}) into eq.~\eqref{A,B,C}, we get
\begin{align}\label{ABC_Nonrelativistic}
\mathcal{A}&=\frac{\lambda^2\alpha}{32\pi}\left[\coth\left(\frac{\pi\alpha}{a}\right)-\frac{4\pi}{\alpha} \mathcal{h}(\alpha, a)\mathcal{u}^2\right]\nonumber\\
\mathcal{B}&=\frac{\lambda^2\alpha}{32\pi}\nonumber\\
\mathcal{C}&=\frac{\lambda^2a}{16\pi^2}\left[1-\left(\frac{7}{6}-\frac{\pi^2}{9}\right)\mathcal{u}^2\right]-\mathcal{A}
\end{align}

\noindent In the limit $\mathcal{u}^2>>1$, the Wightman function can be approximated upto $\mathcal{O}(\frac{1}{\mathcal{u}^4})$ as
\begin{equation}\label{Wightman_UR}
G^{+}(\tau,\tau-\tau')=-\frac{a^2}{16\pi^2}\left[\frac{1}{\sinh^2\left(\frac{a(\tau')}{2\mathcal{u}}-i\epsilon\right)\mathcal{u}^4}\right]\,.
\end{equation}

\noindent Now calculating $\mathcal{G}(\pm \alpha)$ by using the above Wightman function eq.~\eqref{Wightman_UR},
\begin{equation}\label{g1a3}
    \mathcal{G}(\alpha)=\frac{\alpha}{2\pi\mathcal{u}^2}\frac{e^{2\pi\alpha\mathcal{u}/a}}{(e^{2\pi\alpha\mathcal{u}/a}-1)}
\end{equation}
\begin{equation}\label{g1a4}
    \mathcal{G}(-\alpha)=\frac{\alpha}{2\pi\mathcal{u}^2}\frac{1}{(e^{2\pi\alpha\mathcal{u}/a}-1)}
\end{equation}
with, \begin{equation}\label{g102}
\mathcal{G}(0)=\frac{a}{4\pi^2\mathcal{u}^3}~.
\end{equation}
Using eq.(s) (\ref{g1a3}, \ref{g1a4}, \ref{g102}) into eq.~\eqref{A,B,C}, we get
\begin{align}\label{ABC_Ultrarelativistic}
\mathcal{A}&=\frac{\lambda^2\alpha}{32\pi\mathcal{u}^2}\coth\left(\frac{\pi\alpha\mathcal{u}}{a}\right),\hspace{0.5cm}\mathcal{B}=\frac{\lambda^2\alpha}{32\pi\mathcal{u}^2},\hspace{0.5cm}
\mathcal{C}=\frac{\lambda^2a}{16\pi^2\mathcal{u}^3}-\mathcal{A}~.
\end{align}
At this point, it should be noted that in the limit $\mathcal{u}\rightarrow\infty$, all the above constants vanish. Therefore, in this limit the system may be considered as a closed system. Due to the combined effect of acceleration and velocity, this kind of behaviour is shown to
emerge in the earlier works \cite{Abdolrahimi2014, Liu2021}.
        \end{appendices}


\begin{thebibliography}{100}
	\newcommand{\enquote}[1]{``#1''}
	
	\bibitem{kosloff2013quantum}
	R.~Kosloff, \enquote{Quantum thermodynamics: A dynamical viewpoint},
	\href{https://dx.doi.org/10.3390/e15062100}{\emph{Entropy} \textbf{15[6]}
		(2013) 2100}, \href{https://www.mdpi.com/1099-4300/15/6/2100}{{\tt URL}}.
	
	\bibitem{Vinjanampathy2016}
	S.~Vinjanampathy and J.~Anders, \enquote{Quantum thermodynamics},
	\href{https://dx.doi.org/10.1080/00107514.2016.1201896}{\emph{Contemporary
			Physics} \textbf{57[4]} (2016) 545–579}.
	
	\bibitem{Bhattacharjee2021}
	S.~Bhattacharjee and A.~Dutta, \enquote{Quantum thermal machines and
		batteries},
	\href{https://dx.doi.org/10.1140/epjb/s10051-021-00235-3}{\emph{The European
			Physical Journal B} \textbf{94[12]} (2021) 239}.
	
	\bibitem{RMP_QB_2024}
	F.~Campaioli, S.~Gherardini, J.~Q. Quach, M.~Polini and G.~M. Andolina,
	\enquote{Colloquium: Quantum batteries},
	\href{https://dx.doi.org/10.1103/RevModPhys.96.031001}{\emph{Rev. Mod. Phys.}
		\textbf{96} (2024) 031001}.
	
	\bibitem{Alicki2013}
	R.~Alicki and M.~Fannes, \enquote{Entanglement boost for extractable work from
		ensembles of quantum batteries},
	\href{https://dx.doi.org/10.1103/PhysRevE.87.042123}{\emph{Phys. Rev. E}
		\textbf{87} (2013) 042123}.
	
	\bibitem{Campaioli2018}
	F.~Campaioli, F.~A. Pollock and S.~Vinjanampathy,
	\href{https://dx.doi.org/10.1007/978-3-319-99046-0_8}{\enquote{Quantum
			Batteries}, }\emph{in} Thermodynamics in the Quantum Regime: Fundamental
	Aspects and New Directions (edited by F.~Binder, L.~A. Correa, C.~Gogolin,
	J.~Anders and G.~Adesso), p. 207--225, Springer International Publishing,
	Cham 2018.
	
	\bibitem{Campisi2011}
	M.~Campisi, P.~H\"anggi and P.~Talkner, \enquote{Colloquium: Quantum
		fluctuation relations: Foundations and applications},
	\href{https://dx.doi.org/10.1103/RevModPhys.83.771}{\emph{Rev. Mod. Phys.}
		\textbf{83} (2011) 771}.
	
	\bibitem{Horodecki2013}
	M.~Horodecki and J.~Oppenheim, \enquote{Fundamental limitations for quantum and
		nanoscale thermodynamics},
	\href{https://dx.doi.org/10.1038/ncomms3059}{\emph{Nature Communications}
		\textbf{4[1]} (2013) 2059}.
	
	\bibitem{Goold2016}
	J.~Goold, M.~Huber, A.~Riera, L.~del Rio and P.~Skrzypczyk, \enquote{The role
		of quantum information in thermodynamics—a topical review},
	\href{https://dx.doi.org/10.1088/1751-8113/49/14/143001}{\emph{Journal of
			Physics A: Mathematical and Theoretical} \textbf{49[14]} (2016) 143001}.
	
	\bibitem{Binder2015}
	F.~C. Binder, S.~Vinjanampathy, K.~Modi and J.~Goold, \enquote{Quantacell:
		powerful charging of quantum batteries},
	\href{https://dx.doi.org/10.1088/1367-2630/17/7/075015}{\emph{New Journal of
			Physics} \textbf{17[7]} (2015) 075015}.
	
	\bibitem{Binder2017}
	F.~Campaioli, F.~A. Pollock, F.~C. Binder, L.~C\'eleri, J.~Goold,
	S.~Vinjanampathy and K.~Modi, \enquote{Enhancing the Charging Power of
		Quantum Batteries},
	\href{https://dx.doi.org/10.1103/PhysRevLett.118.150601}{\emph{Phys. Rev.
			Lett.} \textbf{118} (2017) 150601}.
	
	\bibitem{Farina2019}
	D.~Farina, G.~M. Andolina, A.~Mari, M.~Polini and V.~Giovannetti,
	\enquote{Charger-mediated energy transfer for quantum batteries: An
		open-system approach},
	\href{https://dx.doi.org/10.1103/PhysRevB.99.035421}{\emph{Phys. Rev. B}
		\textbf{99} (2019) 035421}.
	
	\bibitem{Zhang_YY2019}
	Y.-Y. Zhang, T.-R. Yang, L.~Fu and X.~Wang, \enquote{Powerful harmonic charging
		in a quantum battery},
	\href{https://dx.doi.org/10.1103/PhysRevE.99.052106}{\emph{Phys. Rev. E}
		\textbf{99} (2019) 052106}.
	
	\bibitem{Carrasco2022}
	J.~Carrasco, J.~R. Maze, C.~Hermann-Avigliano and F.~Barra, \enquote{Collective
		enhancement in dissipative quantum batteries},
	\href{https://dx.doi.org/10.1103/PhysRevE.105.064119}{\emph{Phys. Rev. E}
		\textbf{105} (2022) 064119}.
	
	\bibitem{Dario_2022_1}
	J.-Y. Gyhm, D.~\ifmmode~\check{S}\else \v{S}\fi{}afr\'anek and D.~Rosa,
	\enquote{Quantum Charging Advantage Cannot Be Extensive without Global
		Operations},
	\href{https://dx.doi.org/10.1103/PhysRevLett.128.140501}{\emph{Phys. Rev.
			Lett.} \textbf{128} (2022) 140501}.
	
	\bibitem{Rodriguez2023}
	R.~R. Rodr\'{\i}guez, B.~Ahmadi, P.~Mazurek, S.~Barzanjeh, R.~Alicki and
	P.~Horodecki, \enquote{Catalysis in charging quantum batteries},
	\href{https://dx.doi.org/10.1103/PhysRevA.107.042419}{\emph{Phys. Rev. A}
		\textbf{107} (2023) 042419}.
	
	\bibitem{Gemme2023}
	G.~Gemme, G.~M. Andolina, F.~M.~D. Pellegrino, M.~Sassetti and D.~Ferraro,
	\enquote{Off-Resonant Dicke Quantum Battery: Charging by Virtual Photons},
	\href{https://dx.doi.org/10.3390/batteries9040197}{\emph{Batteries}
		\textbf{9[4]}}, \href{https://www.mdpi.com/2313-0105/9/4/197}{{\tt URL}}.
	
	\bibitem{Gemme2024}
	G.~Gemme, M.~Grossi, S.~Vallecorsa, M.~Sassetti and D.~Ferraro, \enquote{Qutrit
		quantum battery: Comparing different charging protocols},
	\href{https://dx.doi.org/10.1103/PhysRevResearch.6.023091}{\emph{Phys. Rev.
			Res.} \textbf{6} (2024) 023091}.
	
	\bibitem{Song2024}
	W.-L. Song, H.-B. Liu, B.~Zhou, W.-L. Yang and J.-H. An, \enquote{Remote
		Charging and Degradation Suppression for the Quantum Battery},
	\href{https://dx.doi.org/10.1103/PhysRevLett.132.090401}{\emph{Phys. Rev.
			Lett.} \textbf{132} (2024) 090401}.
	
	\bibitem{Quach2022}
	J.~Q. Quach, K.~E. McGhee, L.~Ganzer, D.~M. Rouse, B.~W. Lovett, E.~M. Gauger,
	J.~Keeling, G.~Cerullo, D.~G. Lidzey and T.~Virgili, \enquote{Superabsorption
		in an organic microcavity: Toward a quantum battery},
	\href{https://dx.doi.org/10.1126/sciadv.abk3160}{\emph{Science Advances}
		\textbf{8[2]} (2022) eabk3160},
	\href{https://arxiv.org/abs/https://www.science.org/doi/pdf/10.1126\\
		/sciadv.abk3160}{{\tt arXiv:https://www.science.org/doi/pdf/10.1126\\
			/sciadv.abk3160}}.
	
	\bibitem{Gemme2022}
	G.~Gemme, M.~Grossi, D.~Ferraro, S.~Vallecorsa and M.~Sassetti, \enquote{IBM
		Quantum Platforms: A Quantum Battery Perspective},
	\href{https://dx.doi.org/10.3390/batteries8050043}{\emph{Batteries}
		\textbf{8[5]}}, \href{https://www.mdpi.com/2313-0105/8/5/43}{{\tt URL}}.
	
	\bibitem{Hu_CK2022}
	C.-K. Hu, J.~Qiu, P.~J.~P. Souza, J.~Yuan, Y.~Zhou, L.~Zhang, J.~Chu, X.~Pan,
	L.~Hu, J.~Li, Y.~Xu, Y.~Zhong, S.~Liu, F.~Yan, D.~Tan, R.~Bachelard, C.~J.
	Villas-Boas, A.~C. Santos and D.~Yu, \enquote{Optimal charging of a
		superconducting quantum battery},
	\href{https://dx.doi.org/10.1088/2058-9565/ac8444}{\emph{Quantum Science and
			Technology} \textbf{7[4]} (2022) 045018}.
	
	\bibitem{Wenniger2023}
	I.~Maillette~de Buy~Wenniger, S.~E. Thomas, M.~Maffei, S.~C. Wein, M.~Pont,
	N.~Belabas, S.~Prasad, A.~Harouri, A.~Lema\^{\i}tre, I.~Sagnes, N.~Somaschi,
	A.~Auff\`eves and P.~Senellart, \enquote{Experimental Analysis of Energy
		Transfers between a Quantum Emitter and Light Fields},
	\href{https://dx.doi.org/10.1103/PhysRevLett.131.260401}{\emph{Phys. Rev.
			Lett.} \textbf{131} (2023) 260401}.
	
	\bibitem{Lu_2024}
	Z.-G. Lu, G.~Tian, X.-Y. Lü and C.~Shang, \enquote{Topological Quantum
		Batteries},  2024.
	
	\bibitem{Tirone_2024_1}
	S.~Tirone, G.~M. Andolina, G.~Calajò, V.~Giovannetti and D.~Rossini,
	\enquote{Many-body enhancement of energy storage in a waveguide-QED quantum
		battery},  2024.
	
	\bibitem{Dario_2020}
	D.~Rossini, G.~M. Andolina, D.~Rosa, M.~Carrega and M.~Polini, \enquote{Quantum
		Advantage in the Charging Process of Sachdev-Ye-Kitaev Batteries},
	\href{https://dx.doi.org/10.1103/PhysRevLett.125.236402}{\emph{Phys. Rev.
			Lett.} \textbf{125} (2020) 236402}.
	
	\bibitem{Dario_JHEP_2020}
	D.~Rosa, D.~Rossini, G.~M. Andolina, M.~Polini and M.~Carrega,
	\enquote{Ultra-stable charging of fast-scrambling SYK quantum batteries},
	\href{https://dx.doi.org/10.1007/JHEP11(2020)067}{\emph{Journal of High
			Energy Physics} \textbf{2020[11]} (2020) 67}.
	
	\bibitem{Dario_2022}
	J.~Kim, J.~Murugan, J.~Olle and D.~Rosa, \enquote{Operator delocalization in
		quantum networks},
	\href{https://dx.doi.org/10.1103/PhysRevA.105.L010201}{\emph{Phys. Rev. A}
		\textbf{105} (2022) L010201}.
	
	\bibitem{Shaghaghi_2022}
	V.~Shaghaghi, V.~Singh, G.~Benenti and D.~Rosa, \enquote{Micromasers as quantum
		batteries}, \href{https://dx.doi.org/10.1088/2058-9565/ac8829}{\emph{Quantum
			Science and Technology} \textbf{7[4]} (2022) 04LT01}.
	
	\bibitem{Dario_2023}
	C.~Rodr\'{\i}guez, D.~Rosa and J.~Olle, \enquote{Artificial intelligence
		discovery of a charging protocol in a micromaser quantum battery},
	\href{https://dx.doi.org/10.1103/PhysRevA.108.042618}{\emph{Phys. Rev. A}
		\textbf{108} (2023) 042618}.
	
	\bibitem{Carrega2020}
	M.~Carrega, A.~Crescente, D.~Ferraro and M.~Sassetti, \enquote{Dissipative
		dynamics of an open quantum battery},
	\href{https://dx.doi.org/10.1088/1367-2630/abaa01}{\emph{New Journal of
			Physics} \textbf{22[8]} (2020) 083085}.
	
	\bibitem{Tabesh2020}
	F.~T. Tabesh, F.~H. Kamin and S.~Salimi, \enquote{Environment-mediated charging
		process of quantum batteries},
	\href{https://dx.doi.org/10.1103/PhysRevA.102.052223}{\emph{Phys. Rev. A}
		\textbf{102} (2020) 052223}.
	
	\bibitem{Lu_2022}
	C.-Y. Lu, Y.~Cao, C.-Z. Peng and J.-W. Pan, \enquote{Micius quantum experiments
		in space}, \href{https://dx.doi.org/10.1103/RevModPhys.94.035001}{\emph{Rev.
			Mod. Phys.} \textbf{94} (2022) 035001}.
	
	\bibitem{Ribezzo_2023}
	D.~Ribezzo, M.~Zahidy, I.~Vagniluca, N.~Biagi, S.~Francesconi, T.~Occhipinti,
	L.~K. Oxenl{\o}we, M.~Lon{\v{c}}ari{\'{c}}, I.~Cviti{\'{c}},
	M.~Stip{\v{c}}evi{\'{c}}, {\v{Z}}.~Pu{\v{s}}avec, R.~Kaltenbaek,
	A.~Ram{\v{s}}ak, F.~Cesa, G.~Giorgetti, F.~Scazza, A.~Bassi, P.~De~Natale,
	F.~S. Cataliotti, M.~Inguscio, D.~Bacco and A.~Zavatta, \enquote{Deploying an
		Inter-European Quantum Network},
	\href{https://dx.doi.org/10.1002/qute.202200061}{\emph{Advanced Quantum
			Technologies} \textbf{6[2]} (2023) 2200061}.
	
	\bibitem{Fulling}
	S.~A. Fulling, \enquote{Nonuniqueness of Canonical Field Quantization in
		Riemannian Space-Time},
	\href{https://dx.doi.org/10.1103/PhysRevD.7.2850}{\emph{Phys. Rev. D}
		\textbf{7} (1973) 2850}.
	
	\bibitem{Davies1975}
	P.~C.~W. Davies, \enquote{Scalar production in Schwarzschild and Rindler
		metrics}, \href{https://dx.doi.org/10.1088/0305-4470/8/4/022}{\emph{Journal
			of Physics A: Mathematical and General} \textbf{8[4]} (1975) 609}.
	
	\bibitem{Unruh1}
	W.~G. Unruh, \enquote{Notes on black-hole evaporation},
	\href{https://dx.doi.org/10.1103/PhysRevD.14.870}{\emph{Phys. Rev. D}
		\textbf{14} (1976) 870}.
	
	\bibitem{Bruschi2014}
	M.~Ahmadi, D.~E. Bruschi and I.~Fuentes, \enquote{Quantum metrology for
		relativistic quantum fields},
	\href{https://dx.doi.org/10.1103/PhysRevD.89.065028}{\emph{Phys. Rev. D}
		\textbf{89} (2014) 065028}.
	
	\bibitem{Matsas}
	A.~G.~S. Landulfo and G.~E.~A. Matsas, \enquote{Sudden death of entanglement
		and teleportation fidelity loss via the Unruh effect},
	\href{https://dx.doi.org/10.1103/PhysRevA.80.032315}{\emph{Phys. Rev. A}
		\textbf{80} (2009) 032315}.
	
	\bibitem{Y_M_Huang2019}
	Y.~Huang, K.~Yan, Y.~Wu and X.~Hao, \enquote{Decoherence of quantum parameter
		estimation for open Dirac particle in Garfinkle--Horowitz--Strominger
		dilation black hole},
	\href{https://dx.doi.org/10.1140/epjc/s10052-019-7491-z}{\emph{The European
			Physical Journal C} \textbf{79[11]} (2019) 974}.
	
	\bibitem{Z_Zhao2020}
	Z.~Zhao, Q.~Pan and J.~Jing, \enquote{Quantum estimation of acceleration and
		temperature in open quantum system},
	\href{https://dx.doi.org/10.1103/PhysRevD.101.056014}{\emph{Phys. Rev. D}
		\textbf{101} (2020) 056014}.
	
	\bibitem{Du2021}
	H.~Du and R.~B. Mann, \enquote{Fisher information as a probe of spacetime
		structure: relativistic quantum metrology in (A)dS},
	\href{https://dx.doi.org/10.1007/JHEP05(2021)112}{\emph{Journal of High
			Energy Physics} \textbf{2021[5]} (2021) 112}.
	
	\bibitem{Liu2021}
	X.~Liu, J.~Jing, Z.~Tian and W.~Yao, \enquote{Does relativistic motion always
		degrade quantum Fisher information?},
	\href{https://dx.doi.org/10.1103/PhysRevD.103.125025}{\emph{Phys. Rev. D}
		\textbf{103} (2021) 125025}.
	
	\bibitem{Benatti_2004}
	F.~Benatti and R.~Floreanini, \enquote{Entanglement generation in uniformly
		accelerating atoms: Reexamination of the Unruh effect},
	\href{https://dx.doi.org/10.1103/PhysRevA.70.012112}{\emph{Phys. Rev. A}
		\textbf{70} (2004) 012112}.
	
	\bibitem{B_L_Hu2012}
	B.~L. Hu, S.-Y. Lin and J.~Louko, \enquote{Relativistic quantum information in
		detectors–field interactions},
	\href{https://dx.doi.org/10.1088/0264-9381/29/22/224005}{\emph{Classical and
			Quantum Gravity} \textbf{29[22]} (2012) 224005}.
	
	\bibitem{Moustos2017}
	D.~Moustos and C.~Anastopoulos, \enquote{Non-Markovian time evolution of an
		accelerated qubit},
	\href{https://dx.doi.org/10.1103/PhysRevD.95.025020}{\emph{Phys. Rev. D}
		\textbf{95} (2017) 025020}.
	
	\bibitem{Chatterjee2017}
	R.~Chatterjee and A.~S. Majumdar, \enquote{Preservation of quantum coherence
		under Lorentz boost for narrow uncertainty wave packets},
	\href{https://dx.doi.org/10.1103/PhysRevA.96.052301}{\emph{Phys. Rev. A}
		\textbf{96} (2017) 052301}.
	
	\bibitem{Mukherjee_PRA_2024}
	A.~Mukherjee, S.~Sen and S.~Gangopadhyay, \enquote{Quantum coherence measures
		for generalized Gaussian wave packets under a Lorentz boost},
	\href{https://dx.doi.org/10.1103/PhysRevA.110.052413}{\emph{Phys. Rev. A}
		\textbf{110} (2024) 052413}.
	
	\bibitem{Sokolov2020}
	B.~Sokolov, J.~Louko, S.~Maniscalco and I.~Vilja, \enquote{Unruh effect and
		information flow},
	\href{https://dx.doi.org/10.1103/PhysRevD.101.024047}{\emph{Phys. Rev. D}
		\textbf{101} (2020) 024047}.
	
	\bibitem{arias2018unruh}
	E.~Arias, T.~R. de~Oliveira and M.~Sarandy, \enquote{The unruh quantum otto
		engine}, \href{https://dx.doi.org/10.1007/JHEP02(2018)168}{\emph{Journal of
			High Energy Physics} \textbf{2018[2]}}.
	
	\bibitem{Mukherjee2022}
	A.~Mukherjee, S.~Gangopadhyay and A.~S. Majumdar, \enquote{Unruh quantum Otto
		engine in the presence of a reflecting boundary},
	\href{https://dx.doi.org/10.1007/JHEP09(2022)105}{\emph{Journal of High
			Energy Physics} \textbf{2022[9]} (2022) 105}.
	
	\bibitem{Allahverdyan2004}
	A.~E. Allahverdyan, R.~Balian and T.~M. Nieuwenhuizen, \enquote{Maximal work
		extraction from finite quantum systems},
	\href{https://dx.doi.org/10.1209/epl/i2004-10101-2}{\emph{Europhysics
			Letters} \textbf{67[4]} (2004) 565}.
	
	\bibitem{M_N_Bera2020}
	S.~Juli\`a-Farr\'e, T.~Salamon, A.~Riera, M.~N. Bera and M.~Lewenstein,
	\enquote{Bounds on the capacity and power of quantum batteries},
	\href{https://dx.doi.org/10.1103/PhysRevResearch.2.023113}{\emph{Phys. Rev.
			Res.} \textbf{2} (2020) 023113}.
	
	\bibitem{Mir2023}
	X.~Yang, Y.-H. Yang, M.~Alimuddin, R.~Salvia, S.-M. Fei, L.-M. Zhao,
	S.~Nimmrichter and M.-X. Luo, \enquote{Battery Capacity of Energy-Storing
		Quantum Systems},
	\href{https://dx.doi.org/10.1103/PhysRevLett.131.030402}{\emph{Phys. Rev.
			Lett.} \textbf{131} (2023) 030402}.
	
	\bibitem{Mojaveri2023}
	B.~Mojaveri, R.~Jafarzadeh~Bahrbeig, M.~A. Fasihi and S.~Babanzadeh,
	\enquote{Enhancing the direct charging performance of an open quantum battery
		by adjusting its velocity},
	\href{https://dx.doi.org/10.1038/s41598-023-47193-7}{\emph{Scientific
			Reports} \textbf{13[1]} (2023) 19827}.
	
	\bibitem{Fischer_2024}
	J.-Y. Gyhm and U.~R. Fischer, \enquote{Beneficial and detrimental entanglement
		for quantum battery charging},
	\href{https://dx.doi.org/10.1116/5.0184903}{\emph{AVS Quantum Science}
		\textbf{6[1]} (2024) 012001}.
	
	\bibitem{Tirone_2024}
	S.~Tirone, R.~Salvia, S.~Chessa and V.~Giovannetti, \enquote{{Quantum work
			capacitances: Ultimate limits for energy extraction on noisy quantum
			batteries}},
	\href{https://dx.doi.org/10.21468/SciPostPhys.17.2.041}{\emph{SciPost Phys.}
		\textbf{17} (2024) 041},
	\href{https://scipost.org/10.21468/SciPostPhys.17.2.041}{{\tt URL}}.
	
	\bibitem{Hao2023}
	X.~Hao, K.~Yan, J.~Tan and Q.-Y. Wu, \enquote{Quantum work extraction of an
		accelerated Unruh-DeWitt battery in relativistic motion},
	\href{https://dx.doi.org/10.1103/PhysRevA.107.012207}{\emph{Phys. Rev. A}
		\textbf{107} (2023) 012207}.
	
	\bibitem{Abdolrahimi2014}
	S.~Abdolrahimi, \enquote{Velocity effects on an accelerated Unruh–DeWitt
		detector},
	\href{https://dx.doi.org/10.1088/0264-9381/31/13/135009}{\emph{Classical and
			Quantum Gravity} \textbf{31[13]} (2014) 135009}.
	
	\bibitem{Tian_2024}
	Z.~Tian, X.~Liu, J.~Wang and J.~Jing, \enquote{Dissipative dynamics of an open
		quantum battery in the BTZ spacetime},  2024.
	
	\bibitem{DeWitt1979}
	B.~DeWitt, General Relativity: An Einstein Centenary Survey, Cambridge
	University Press, Cambridge, U.K. 1979.
	
	\bibitem{Unruh1984}
	W.~G. Unruh and R.~M. Wald, \enquote{What happens when an accelerating observer
		detects a Rindler particle},
	\href{https://dx.doi.org/10.1103/PhysRevD.29.1047}{\emph{Phys. Rev. D}
		\textbf{29} (1984) 1047}.
	
	\bibitem{hu2012geometric}
	J.~Hu and H.~Yu, \enquote{Geometric phase for an accelerated two-level atom and
		the Unruh effect},
	\href{https://dx.doi.org/10.1103/PhysRevA.85.032105}{\emph{Phys. Rev. A}
		\textbf{85} (2012) 032105}.
	
	\bibitem{Martinez2016_1}
	A.~Pozas-Kerstjens and E.~Mart\'{\i}n-Mart\'{\i}nez, \enquote{Entanglement
		harvesting from the electromagnetic vacuum with hydrogenlike atoms},
	\href{https://dx.doi.org/10.1103/PhysRevD.94.064074}{\emph{Phys. Rev. D}
		\textbf{94} (2016) 064074}.
	
	\bibitem{yu2008understanding}
	H.~Yu and J.~Zhang, \enquote{Understanding Hawking radiation in the framework
		of open quantum systems},
	\href{https://dx.doi.org/10.1103/PhysRevD.77.024031}{\emph{Phys. Rev. D}
		\textbf{77} (2008) 024031}.
	
	\bibitem{jin2014dynamical}
	Y.~Jin, J.~Hu and H.~Yu, \enquote{Dynamical behavior and geometric phase for a
		circularly accelerated two-level atom},
	\href{https://dx.doi.org/10.1103/PhysRevA.89.064101}{\emph{Phys. Rev. A}
		\textbf{89} (2014) 064101}.
	
	\bibitem{liu2021relativistic}
	X.~Liu, J.~Jing, Z.~Tian and W.~Yao, \enquote{Does relativistic motion always
		degrade quantum Fisher information?},
	\href{https://dx.doi.org/10.1103/PhysRevD.103.125025}{\emph{Phys. Rev. D}
		\textbf{103} (2021) 125025}.
	
	\bibitem{zhang2007unruh}
	J.~Zhang and H.~Yu, \enquote{Unruh effect and entanglement generation for
		accelerated atoms near a reflecting boundary},
	\href{https://dx.doi.org/10.1103/PhysRevD.75.104014}{\emph{Phys. Rev. D}
		\textbf{75} (2007) 104014}.
	
	\bibitem{hu2015entanglement}
	J.~Hu and H.~Yu, \enquote{Entanglement dynamics for uniformly accelerated
		two-level atoms},
	\href{https://dx.doi.org/10.1103/PhysRevA.91.012327}{\emph{Phys. Rev. A}
		\textbf{91} (2015) 012327}.
	
	\bibitem{yang2016entanglement}
	Y.~Yang, J.~Hu and H.~Yu, \enquote{Entanglement dynamics for uniformly
		accelerated two-level atoms coupled with electromagnetic vacuum
		fluctuations},
	\href{https://dx.doi.org/10.1103/PhysRevA.94.032337}{\emph{Phys. Rev. A}
		\textbf{94} (2016) 032337}.
	
	\bibitem{cheng2018entanglement}
	S.~Cheng, H.~Yu and J.~Hu, \enquote{Entanglement dynamics for uniformly
		accelerated two-level atoms in the presence of a reflecting boundary},
	\href{https://dx.doi.org/10.1103/PhysRevD.98.025001}{\emph{Phys. Rev. D}
		\textbf{98} (2018) 025001}.
	
	\bibitem{she2019entanglement}
	J.~She, J.~Hu and H.~Yu, \enquote{Entanglement dynamics for circularly
		accelerated two-level atoms coupled with electromagnetic vacuum
		fluctuations},
	\href{https://dx.doi.org/10.1103/PhysRevD.99.105009}{\emph{Phys. Rev. D}
		\textbf{99} (2019) 105009}.
	
	\bibitem{zhou2021entanglement}
	Y.~Zhou, J.~Hu and H.~Yu, \enquote{Entanglement dynamics for Unruh-DeWitt
		detectors interacting with massive scalar fields: the Unruh and anti-Unruh
		effects}, \href{https://dx.doi.org/10.1007/JHEP09(2021)088}{\emph{Journal of
			High Energy Physics} \textbf{2021[9]} (2021) 88}.
	
	\bibitem{chen2022entanglement}
	Y.~Chen, J.~Hu and H.~Yu, \enquote{Entanglement generation for uniformly
		accelerated atoms assisted by environment-induced interatomic interaction and
		the loss of the anti-Unruh effect},
	\href{https://dx.doi.org/10.1103/PhysRevD.105.045013}{\emph{Phys. Rev. D}
		\textbf{105} (2022) 045013}.
	
	\bibitem{gilles1993tp}
	L.~Gilles and P.~L. Knight, \enquote{Two-photon absorption and nonclassical
		states of light},
	\href{https://dx.doi.org/10.1103/PhysRevA.48.1582}{\emph{Phys. Rev. A}
		\textbf{48} (1993) 1582}.
	
	\bibitem{Wu2023}
	D.~Wu, S.-C. Tang and Y.~Shi, \enquote{Birth and death of entanglement between
		two accelerating Unruh-DeWitt detectors coupled with a scalar field},
	\href{https://dx.doi.org/10.1007/JHEP12(2023)037}{\emph{Journal of High
			Energy Physics} \textbf{2023[12]} (2023) 37}.
	
	\bibitem{hinton1984particle}
	K.~J. Hinton, \enquote{Particle detector equivalence},
	\href{https://dx.doi.org/10.1088/0264-9381/1/1/006}{\emph{Classical and
			Quantum Gravity} \textbf{1[1]} (1984) 27}.
	
	\bibitem{Hummer2016}
	D.~H\"ummer, E.~Mart\'{\i}n-Mart\'{\i}nez and A.~Kempf, \enquote{Renormalized
		Unruh-DeWitt particle detector models for boson and fermion fields},
	\href{https://dx.doi.org/10.1103/PhysRevD.93.024019}{\emph{Phys. Rev. D}
		\textbf{93} (2016) 024019}.
	
	\bibitem{Sachs2017}
	A.~Sachs, R.~B. Mann and E.~Mart\'{\i}n-Mart\'{\i}nez, \enquote{Entanglement
		harvesting and divergences in quadratic Unruh-DeWitt detector pairs},
	\href{https://dx.doi.org/10.1103/PhysRevD.96.085012}{\emph{Phys. Rev. D}
		\textbf{96} (2017) 085012}.
	
	\bibitem{Gray2018}
	F.~Gray and R.~B. Mann, \enquote{Scalar and fermionic Unruh Otto engines},
	\href{https://dx.doi.org/10.1007/JHEP11(2018)174}{\emph{Journal of High
			Energy Physics} \textbf{2018[11]} (2018) 174}.
	
	\bibitem{Koch_2016}
	C.~P. Koch, \enquote{Controlling open quantum systems: tools, achievements, and
		limitations},
	\href{https://dx.doi.org/10.1088/0953-8984/28/21/213001}{\emph{Journal of
			Physics: Condensed Matter} \textbf{28[21]} (2016) 213001}.
	
	\bibitem{Uchiyama_2018}
	C.~Uchiyama, W.~J. Munro and K.~Nemoto, \enquote{Environmental engineering for
		quantum energy transport},
	\href{https://dx.doi.org/10.1038/s41534-018-0079-x}{\emph{npj Quantum
			Information} \textbf{4[1]} (2018) 33}.
	
	\bibitem{munoz2021quantum}
	C.~S\'anchez Mu\~noz, G.~Frascella and F.~Schlawin, \enquote{Quantum metrology
		of two-photon absorption},
	\href{https://dx.doi.org/10.1103/PhysRevResearch.3.033250}{\emph{Phys. Rev.
			Res.} \textbf{3} (2021) 033250}.
	
	\bibitem{alushi2023waveguide}
	U.~Alushi, T.~Ramos, J.~J. Garc\'{\i}a-Ripoll, R.~Di~Candia and S.~Felicetti,
	\enquote{Waveguide QED with Quadratic Light-Matter Interactions},
	\href{https://dx.doi.org/10.1103/PRXQuantum.4.030326}{\emph{PRX Quantum}
		\textbf{4} (2023) 030326}.
	
	\bibitem{felicetti2018tp}
	S.~Felicetti, D.~Z. Rossatto, E.~Rico, E.~Solano and P.~Forn-D\'{\i}az,
	\enquote{Two-photon quantum Rabi model with superconducting circuits},
	\href{https://dx.doi.org/10.1103/PhysRevA.97.013851}{\emph{Phys. Rev. A}
		\textbf{97} (2018) 013851}.
	
	\bibitem{felicetti2018ultra}
	S.~Felicetti, M.-J. Hwang and A.~Le~Boit\'e, \enquote{Ultrastrong-coupling
		regime of nondipolar light-matter interactions},
	\href{https://dx.doi.org/10.1103/PhysRevA.98.053859}{\emph{Phys. Rev. A}
		\textbf{98} (2018) 053859}.
	
	\bibitem{zhou2006nonlinear}
	X.~Zhou and A.~Mizel, \enquote{Nonlinear Coupling of Nanomechanical Resonators
		to Josephson Quantum Circuits},
	\href{https://dx.doi.org/10.1103/PhysRevLett.97.267201}{\emph{Phys. Rev.
			Lett.} \textbf{97} (2006) 267201}.
	
	\bibitem{scully1997quantum}
	M.~O. Scully and M.~S. Zubairy,
	\href{https://dx.doi.org/10.1017/CBO9780511813993}{Quantum optics}, Cambridge
	university press 1997,
	\href{https://www.cambridge.org/core/books/quantum-optics/08DC53888452CBC6CDC0FD8A1A1A4DD7}{{\tt
			URL}}.
	
	\bibitem{GKS_1976}
	V.~Gorini, A.~Kossakowski and E.~C.~G. Sudarshan, \enquote{{Completely positive
			dynamical semigroups of N‐level systems}},
	\href{https://dx.doi.org/10.1063/1.522979}{\emph{Journal of Mathematical
			Physics} \textbf{17[5]} (1976) 821},
	\href{https://arxiv.org/abs/https://pubs.aip.org/aip/jmp/article-pdf\\/17/5/821/19090720/821\_1\_online.pdf}{{\tt
			arXiv:https://pubs.aip.org/aip/jmp/article-pdf\\/17/5/821/19090720/821\_1\_online.pdf}}.
	
	\bibitem{Lindblad1976}
	G.~Lindblad, \enquote{On the generators of quantum dynamical semigroups},
	\href{https://dx.doi.org/10.1007/BF01608499}{\emph{Communications in
			Mathematical Physics} \textbf{48[2]} (1976) 119}.
	
	\bibitem{birrell1984quantum}
	N.~D. Birrell and P.~Davies,
	\href{https://dx.doi.org/10.1017/CBO9780511622632}{Quantum fields in curved
		space}, Cambridge university press 1984,
	\href{https://www.cambridge.org/core/books/quantum-fields-in-curved-space/95376B0CAD78EE767FCD6205F8327F4C}{{\tt
			URL}}.
	
	\bibitem{Chatterjee2021_1}
	R.~Chatterjee, S.~Gangopadhyay and A.~S. Majumdar, \enquote{Violation of
		equivalence in an accelerating atom-mirror system in the generalized
		uncertainty principle framework},
	\href{https://dx.doi.org/10.1103/PhysRevD.104.124001}{\emph{Phys. Rev. D}
		\textbf{104} (2021) 124001}.
	
	\bibitem{Mukherjee2023}
	A.~Mukherjee, S.~Gangopadhyay and A.~S. Majumdar, \enquote{Fulling-Davies-Unruh
		effect for accelerated two-level single and entangled atomic systems},
	\href{https://dx.doi.org/10.1103/PhysRevD.108.085018}{\emph{Phys. Rev. D}
		\textbf{108} (2023) 085018}.
	
	\bibitem{carmichael1999statistical}
	H.~J. Carmichael,
	\href{https://dx.doi.org/10.1007/978-3-540-47620-7}{Statistical Methods in
		Quantum Optics 1: Master Equations and Fokker-Planck Equations}, Theoretical
	and Mathematical Physics, Springer 1999.
	
	\bibitem{Breuer2007}
	H.-P. Breuer and F.~Petruccione,
	\href{https://dx.doi.org/10.1093/acprof:oso/9780199213900.001.0001}{{The
			Theory of Open Quantum Systems}}, Oxford University Press 2007.
	
	\bibitem{Yao_2015}
	Y.~Jin and H.~Yu, \enquote{Electromagnetic shielding in quantum metrology},
	\href{https://dx.doi.org/10.1103/PhysRevA.91.022120}{\emph{Phys. Rev. A}
		\textbf{91} (2015) 022120}.
	
	\bibitem{Myers_2022}
	N.~M. Myers, O.~Abah and S.~Deffner, \enquote{{Quantum thermodynamic devices:
			From theoretical proposals to experimental reality}},
	\href{https://dx.doi.org/10.1116/5.0083192}{\emph{AVS Quantum Science}
		\textbf{4[2]} (2022) 027101},
	\href{https://arxiv.org/abs/https://pubs.aip.org/avs/aqs/article-pdf/doi/\\10.1116/5.0083192/16494008/027101\_1\_online.pdf}{{\tt
			arXiv:https://pubs.aip.org/avs/aqs/article-pdf/doi/\\10.1116/5.0083192/16494008/027101\_1\_online.pdf}}.
	
	\bibitem{ann2022tp}
	B.-m. Ann, W.~Kessels and G.~A. Steele, \enquote{Two-photon sideband
		interaction in a driven quantum Rabi model: Quantitative discussions with
		derived longitudinal drives and beyond the rotating wave approximation},
	\href{https://dx.doi.org/10.1103/PhysRevResearch.4.013005}{\emph{Phys. Rev.
			Res.} \textbf{4} (2022) 013005}.
	
	\bibitem{wang2016method}
	X.~Wang, A.~Miranowicz, H.-R. Li and F.~Nori, \enquote{Method for observing
		robust and tunable phonon blockade in a nanomechanical resonator coupled to a
		charge qubit},
	\href{https://dx.doi.org/10.1103/PhysRevA.93.063861}{\emph{Phys. Rev. A}
		\textbf{93} (2016) 063861}.
	
	\bibitem{munoz2018hybrid}
	C.~S\'anchez Mu\~noz, A.~Lara, J.~Puebla and F.~Nori, \enquote{Hybrid Systems
		for the Generation of Nonclassical Mechanical States via Quadratic
		Interactions},
	\href{https://dx.doi.org/10.1103/PhysRevLett.121.123604}{\emph{Phys. Rev.
			Lett.} \textbf{121} (2018) 123604}.
	
	\bibitem{felicetti2015spectral}
	S.~Felicetti, J.~S. Pedernales, I.~L. Egusquiza, G.~Romero, L.~Lamata, D.~Braak
	and E.~Solano, \enquote{Spectral collapse via two-phonon interactions in
		trapped ions},
	\href{https://dx.doi.org/10.1103/PhysRevA.92.033817}{\emph{Phys. Rev. A}
		\textbf{92} (2015) 033817}.
	
	\bibitem{cheng2018nonlinear}
	X.-H. Cheng, I.~n. Arrazola, J.~S. Pedernales, L.~Lamata, X.~Chen and
	E.~Solano, \enquote{Nonlinear quantum Rabi model in trapped ions},
	\href{https://dx.doi.org/10.1103/PhysRevA.97.023624}{\emph{Phys. Rev. A}
		\textbf{97} (2018) 023624}.
	
	\bibitem{puebla2019quantum}
	R.~Puebla, J.~Casanova, O.~Houhou, E.~Solano and M.~Paternostro,
	\enquote{Quantum simulation of multiphoton and nonlinear dissipative
		spin-boson models},
	\href{https://dx.doi.org/10.1103/PhysRevA.99.032303}{\emph{Phys. Rev. A}
		\textbf{99} (2019) 032303}.
	
	\bibitem{cong2020selective}
	L.~Cong, S.~Felicetti, J.~Casanova, L.~Lamata, E.~Solano and I.~Arrazola,
	\enquote{Selective interactions in the quantum Rabi model},
	\href{https://dx.doi.org/10.1103/PhysRevA.101.032350}{\emph{Phys. Rev. A}
		\textbf{101} (2020) 032350}.
	
	\bibitem{schneeweiss2018cold}
	P.~Schneeweiss, A.~Dareau and C.~Sayrin, \enquote{Cold-atom-based
		implementation of the quantum Rabi model},
	\href{https://dx.doi.org/10.1103/PhysRevA.98.021801}{\emph{Phys. Rev. A}
		\textbf{98} (2018) 021801}.
	
	\bibitem{dareau2018observation}
	A.~Dareau, Y.~Meng, P.~Schneeweiss and A.~Rauschenbeutel, \enquote{Observation
		of Ultrastrong Spin-Motion Coupling for Cold Atoms in Optical Microtraps},
	\href{https://dx.doi.org/10.1103/PhysRevLett.121.253603}{\emph{Phys. Rev.
			Lett.} \textbf{121} (2018) 253603}.
	
	\bibitem{goetz2018parity}
	J.~Goetz, F.~Deppe, K.~G. Fedorov, P.~Eder, M.~Fischer, S.~Pogorzalek, E.~Xie,
	A.~Marx and R.~Gross, \enquote{Parity-Engineered Light-Matter Interaction},
	\href{https://dx.doi.org/10.1103/PhysRevLett.121.060503}{\emph{Phys. Rev.
			Lett.} \textbf{121} (2018) 060503}.
	
\end{thebibliography}
\end{document}